\documentclass[a4paper, 9pt,twocolumn]{article}

\usepackage{geometry}                
\usepackage{amsmath} 
\usepackage{amssymb}  
\usepackage{mathtools}
\usepackage{physics}
\usepackage{authblk}
\usepackage[colorlinks=true,linkcolor=blue,citecolor=black]{hyperref}
\usepackage[font=footnotesize,labelfont=bf]{caption} 
\usepackage{verbatim} 
\usepackage{tabularx}

\def\Nat{\mathbb{N}_0}       
\def\SpaceX{\mathbb{X}}    
\def\n{n}                               
\def\vn{\mathbf{\n}}                   
\newcommand{\nx}[1]{\n\pqty{#1}}     
\def\P{P}                            
\newcommand{\Et}[1]{\langle #1 \rangle}
\newcommand{\Etb}[1]{\Big\langle #1 \Big\rangle}
\def\Mom{M} 
\newcommand{\Momx}[1]{\Mom^{#1}}

\def\ReactantS{a}
\def\ReactantSv{\mathbf{\ReactantS}}
\def\ProductS{b}
\def\ProductSv{\mathbf{\ProductS}}
\def\NumReact{r}
\def\NumProd{p}

\def\Cset{ \mathcal{C}}
\def\Jset{ \mathcal{J}}
\def\xc{\mathbf{X}}
\def\yc{\mathbf{Y}}

\def\prop{h}              
\def\Prop{H}              

\def\d{\mathrm{d}}

\begin{document}

\title{\LARGE \bf
Stochastic reaction networks in dynamic compartment populations
}

\author[1,2]{Lorenzo Duso}
\author[1,2,3]{Christoph Zechner}

\affil[1]{Max Planck Institute of Molecular Cell Biology and Genetics, 01307 Dresden, Germany}
\affil[2]{Center for Systems Biology Dresden, 01307 Dresden, Germany}
\affil[3]{Correspondence to: zechner@mpi-cbg.de}
\date{}
\maketitle


{\noindent \fontsize{9}{0} \selectfont \textbf{
Abstract -- Compartmentalization of biochemical processes underlies all biological systems, from the organelle to the tissue scale. Theoretical models to study the interplay between noisy reaction dynamics and compartmentalization are sparse, and typically very challenging to analyze computationally. Recent studies have made progress towards addressing this problem in the context of concrete biological systems but general approaches remain lacking. 
In this work we propose a mathematical framework based on counting processes that allows us to study compartment populations with arbitrary interactions and internal biochemistry. We provide an efficient description of the population dynamics in terms of differential equations which capture moments of the population and their variability. 
We demonstrate the relevance of our approach using several case studies inspired by biological systems at different scales.
}}


\section{Introduction}

Compartmentalization is inherent to all forms of life~\cite{Alberts2002}.
By separating biochemical processes from their surroundings, compartments serve as spatial and functional building blocks that govern biological organization at different scales. 
At the subcellular level, for instance, networks of vesicles collectively regulate the delivery, sorting and breakdown of molecular cargo~\cite{Doherty2009,Villasenor2016}.
At the tissue scale, cells themselves act as functional units, each executing its internal biochemical program while interacting with the surrounding cell population and environment. 
These and other forms of compartmentalization have in common that an emergent behavior or function is achieved through the collective dynamics of multiple interacting compartments, which are in complex interplay with their environment as well as the biochemical processes they carry.

In many biological systems, the compartments as well as their molecular contents are present in low copy numbers, such that random fluctuations in their dynamics become important \cite{Mcadams1997,Bialek2012}. 
In certain cases, fluctuations are even exploited to generate heterogeneity among a population \cite{Elowitz2010}. 
Thus, investigating the dynamical properties of compartmentalized systems, especially in the presence of random fluctuations, is an important task towards understanding living systems across different scales.

From a methodological perspective, stochasticity poses formidable challenges in the analysis of biochemical processes in the presence of dynamic compartmentalization.
Even in absence of the latter, the treatment of stochastic reaction dynamics is demanding and a large body of literature from the past decades has been devoted to addressing this subject~\cite{Gillespie2007,munsky2006finite,Schnoerr2017}.
Among the few available approaches to combine stochastic reaction- and compartmental-dynamics, population balance equations (PBEs) are among the most prominent~\cite{Henson2003,Ramkrishna2014, Waldherr2018}.
In this context, PBEs describe the time evolution of the number density of a compartment population due to compartment interactions or internal compartment dynamics, which may stem from chemical reactions or material exchange. 
Despite their popularity and numerous applications across various fields of science~\cite{Ramkrishna2000, Krapivsky2010, Jacobson2005}, PBEs are most commonly found as integro-partial differential equations in mean-field form: rather than the actual number density, they describe an expected number density, often motivated using scaling arguments~\cite{Ramkrishna1973}. Therefore, information about mesoscopic fluctuations is necessarily lost.

The relevance of noise in biological systems has led to an increased interest in stochastic population balance modeling~\cite{Ramkrishna2014,Shu2011}. A few recent elegant studies, for instance, show how cell proliferation can be coupled with stochastic cell-internal dynamics~\cite{Thomas2019, Wu2013_NanogPBE}. 
Since the adoption of a stochastic number density severely complicates the mathematical treatment, results can be often achieved only by imposing tailored approximations, simplifying assumptions, or using costly forward Monte Carlo simulation.
In summary, while stochastic population balance has generated important insights into different biological applications, a comprehensive computational framework remains lacking. 

\begin{figure}[t]
\centering
\includegraphics[width=.9\linewidth]{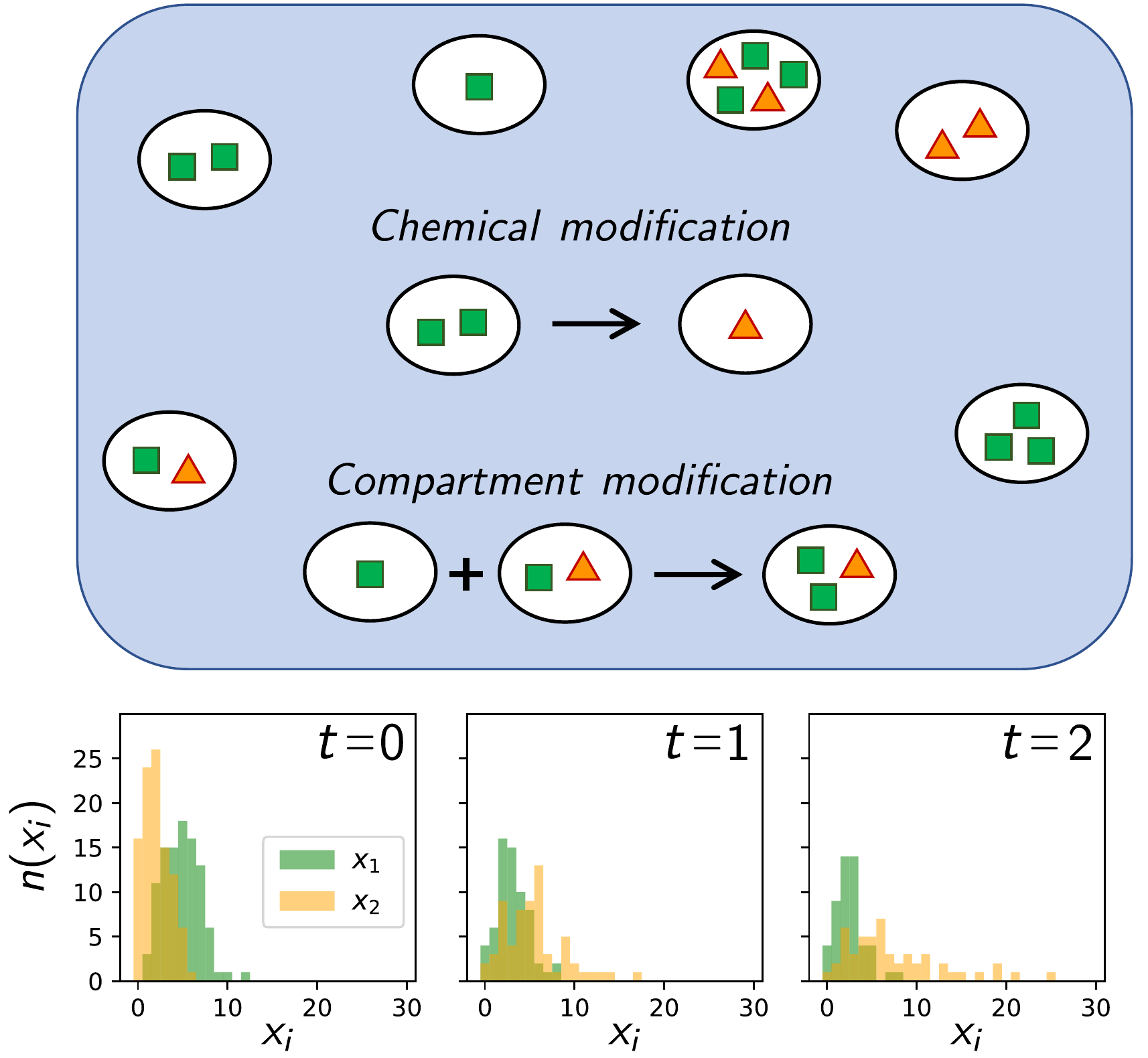}
\caption{Schematic illustration of a compartment population exhibiting chemical and compartmental dynamics. The bottom panels show the marginals of the joint number distribution $\nx{x_1,x_2}$ at three time points along a stochastic realization.}
\label{fig:ConceptFigure}
\end{figure}

The goal of the present work is to develop a general and computationally efficient approach to study stochastic biochemical processes in populations of dynamically interacting compartments. 
In particular, we consider both the compartments and the molecules inside them as discrete objects that can undergo stochastic events.
Our approach shares similarities with population balance modeling, although we adopt an orthogonal strategy for its analytical and computational treatment.
In analogy to reaction networks, we describe changes in the state of a compartment population using a set of compartment-stoichiometric equations.
This allows us to model arbitrary interactions among distinct compartments but also to account for molecular modifications, as illustrated in Fig.~\ref{fig:ConceptFigure}.
Using a counting-process formalism, we introduce an exact stochastic equation that captures how a finite-size population of compartments and their molecular contents evolves with time. 
Realizations of the resulting equation can be efficiently simulated using stochastic simulation algorithms~\cite{gillespie1976general}. 
However, forward simulation can quickly become computationally expensive, and moreover, does not provide analytical insights into the emergent behavior of a population. 
To address this problem, we show how the population dynamics can be expressed compactly in terms of population moments. 
Thanks to the counting-process formalism, complicated manipulations of the number density function can be bypassed, thereby avoiding the difficulties typically encountered using conventional population balance methodology. 
The obtained moment dynamics are themselves stochastic and therefore carry information not only about the average behavior of the population, but also its variability, in contrast to mean-field approaches.   
Using moment-closure approximations, we derive a set of ordinary differential equations, which reveal means and variances of these population moments in a very efficient manner. 
We demonstrate our approach using several case studies inspired by biological systems of different complexity.

\section{THEORETICAL RESULTS}

\subsection{Stochastic compartment populations}
We define a compartment population as a collection of $N$ distinct entities, each being associated with its own molecular state. The state of compartment $i$ is described by a discrete-valued, $D$-dimensional variable $\vb{x}_i = (x_{1, i}, \ldots, x_{D, i}) \in \SpaceX \subseteq \Nat^D$. A single state variable $x_{d, i}$ typically represents the copy number of a particular chemical species present in compartment $i$, but may also be used to capture more coarse-grained compartment attributes such as cell types or vesicle identities. We consider the case where the population and their compartments are characterized by their molecular state only, while other physical properties such as location in space or shape are not taken into account. 
Thus, two compartments with the same molecular state $\vb{x}$ are identical and indistinguishable in our formalism. Correspondingly, we can represent the state of the population by a number distribution function $\nx{\vb{x}} \in \Nat$, which counts the number of compartments in the population that have content equal to $\vb{x}$. 
Furthermore, we define the compartment number array $\vn=(n_{\vb{x}})_{\vb{x} \in\mathcal{\SpaceX}}$ with $n_{\vb{x}} = n(\vb{x})$, which enumerates all compartment numbers within a single (and typically infinitely sized) structure of rank $D$. This array can be understood as a multidimensional matrix, which is used purely for notational convenience. 

We next allow the compartment population to exhibit temporal dynamics. On the one hand, changes in the population may occur because compartments themselves undergo modifications and interact with one another. For instance, two compartments may fuse or a compartment may exit the system. On the other hand, a compartment's state may change due to chemical reactions among its molecules. Regardless of their specific nature, all chemical or compartment modifications can be expressed in terms of changes in the number compartment distribution $n(\vb{x})$.
Formally, we can describe those changes using stoichiometric equations of the form
\begin{equation}
    \sum_{\vb{x} \in \SpaceX} \ReactantS_{j,\vb{x}} [\vb{x}] \xrightharpoonup{~~} \sum_{\vb{x} \in \SpaceX}\ProductS_{j,\vb{x}} [\vb{x}]
\label{eq:ArbitraryPopulationTransition}
\end{equation}
where the symbol $[\vb{x}]$ denotes a compartment of content $\vb{x}$ and the non-negative integers $\ReactantS_{j,\vb{x}}$ and $\ProductS_{j,\vb{x}}$ are the stoichiometric reactant and product coefficients of transition $j$. Furthermore, we define the arrays $\ReactantSv_j = (a_{j, \vb{x}})_{\vb{x} \in \SpaceX}$ and $\ProductSv_j=(b_{j, \vb{x}})_{\vb{x} \in \SpaceX}$, such that the population state $\vn$ changes by $\Delta\vn_j=\ProductSv_j-\ReactantSv_j $ whenever transition $j$ occurs. 
We let $\Jset$ denote the set of all considered transitions.
Using these definitions, we can express the state of the population at any time $t>0$ as
\begin{equation}
	\vn = \vn(0) + \sum_{j \in \Jset} \Delta\vn_j R_j,
\label{eq:RTMsimple}
\end{equation}
with $R_j(t)$ as a counting process that counts the number of occurrences of transition $j$ up to time $t$ and $\vn(0)$ as the initial configuration of the system. Note that time-dependencies are not made explicit in our notation for compactness, but the reader should keep in mind that both $\vn(t)$ and $R_j(t)$ vary with time.

We next equip the counting processes $R_j(t)$ with instantaneous rate functions $\prop_j(t)$ for $j\in \Jset$, which govern how likely each compartment transition happens within an infinitesimal interval of time $(t,t+\d t]$. Throughout this work we consider the rate functions to depend only on the current configuration of the population $\vn(t)$, consistent with Markovian dynamics.
It can then be shown that the counting processes $R_j(t)$ can be expressed as independent, time-transformed unit Poisson processes $\tilde{R}_j(t)$ such that
\begin{equation}
	\vn = \vn(0) + \sum_{j \in \Jset} \Delta\vn_j \tilde{R}_j\pqty{\int_0^t \prop_j(\vn(s)) \d s}.
\label{eq:RTMpopulation}
\end{equation}
Eq. \ref{eq:RTMpopulation} is known as the random time-change representation \cite{Kurtz2011}. 
We can also write the stochastic evolution~\ref{eq:RTMsimple}  of the population state in differential form as
\begin{equation}
	\d\vn = \sum_{j \in \Jset} \Delta\vn_j \d R_j ,
\label{eq:dn}
\end{equation}
where $\d R_j$ is the differential of the counting process $R_j$, and takes value $1$ whenever a transition of type $j$ occurs at time $t$ and is zero otherwise. 

\subsection{Transition Classes}

\begin{table*}[t]
  \centering
    \begin{tabularx}{\textwidth} { 
    |p{\dimexpr.25\linewidth-2\tabcolsep-1.3333\arrayrulewidth}
    p{\dimexpr.25\linewidth-2\tabcolsep-1.3333\arrayrulewidth}
    p{\dimexpr.5 \linewidth-2\tabcolsep-1.3333\arrayrulewidth}|
       }
     \hline
     \textbf{Description} & \textbf{Stoichiometry} & \textbf{Propensity Function} \\
     \hline
     Compartment Intake  & $ \emptyset \xrightharpoonup{\prop_{I}(\vn;\vb{y})} [\vb{y}]$  & $\prop_{I}(\vn;\vb{y})=k_I \pi_I(\vb{y})$  \\
     \hline
     Compartment Exit  & $[\vb{x}] \xrightharpoonup{\prop_{E}(\vn;\vb{x})} \emptyset$  & $\prop_{E}(\vn;\vb{x})=k_E g_E(\vb{x})\nx{\vb{x}}$  \\
     \hline
     Binary Coagulation  & $ [\vb{x}] + [\vb{x}'] \xrightharpoonup{\prop_{C}(\vn;\vb{x},\vb{x}',\vb{y})} [\vb{y}]$  & $\prop_{C}(\vn;\vb{x},\vb{x}',\vb{y}) = k_C g_C(\vb{x},\vb{x}')\frac{n(\vb{x})(n(\vb{x}')-\delta_{\vb{x},\vb{x}'})}{1+\delta_{\vb{x},\vb{x}'}} \delta_{\vb{y},\vb{x}+\vb{x}'}$ \\
     \hline
      Binary Fragmentation  & $ [\vb{x}] \xrightharpoonup{\prop_{F}(\vn;\vb{x},\vb{y},\vb{y}')} [\vb{y}] + [\vb{y}']$  & $\prop_{F}(\vn;\vb{x},\vb{y},\vb{y}')=k_F g_F(\vb{x}) \nx{\vb{x}}\pi_F(\vb{y}|\vb{x})\delta_{\vb{y}',\vb{x}-\vb{y}} $ \\
    \hline
     Chemical Reaction  & $ [\vb{x}] \xrightharpoonup{\prop_{l}(\vn;\vb{x},\vb{y})} [\vb{y}]$  & $\prop_{l}(\vn;\vb{x},\vb{y})= k_l g_l(\vb{x}) \nx{\vb{x}} \delta_{\vb{y},\vb{x}+\Delta\vb{x}_l}$ \\
    \hline
    \end{tabularx}
   \caption{Several examples of population transition classes and the structure of their rate laws. 
  }
  \label{tab:TransitionExamples}
\end{table*}

Eq.~\ref{eq:dn} represents a continuous-time Markov chain formalism for stochastic compartment populations whose dynamics are governed by an arbitrary set of transitions $\Jset$. 
While this representation is very general, it is rather impractical, because in most relevant situations the set $\Jset$ encompasses infinitely many transitions.  
For instance, if two compartments of arbitrary size can fuse with each other, then an infinite number of transitions has to be introduced to model the interaction of all possible pairs of compartment contents.
To address this problem, we introduce a specification of the transitions in terms of a finite set of \textit{transition classes}, which represent generic rules by which compartments of arbitrary content can interact. 
In the case of compartment fusion, for instance, we could define a single transition class which transforms two compartments with content $\vb{x}$ and $\vb{x}'$ into a single compartment with content $\vb{x}+\vb{x}'$, regardless of the specific value of $\vb{x}$ and $\vb{x}'$. 

Formally, we define a transition class in two steps. 
First, we specify the general structure of a transition class $c$ by fixing the number of reactant compartments $\NumReact_c$ and the number of product compartments $\NumProd_c$ that are involved. 
For instance, in the case of compartment fusion we would have $\NumReact_c = 2$ and $\NumProd_c=1$. Throughout this work, we will restrict ourselves to the case $\NumReact_c, \NumProd_c \in \{0,1,2 \}$, but the following discussion holds true also for transitions involving more than two reactant or product compartments. 
We denote with $\xc_c \in \SpaceX^{\NumReact_c}$ and $\yc_c \in \SpaceX^{\NumProd_c}$ the particular choice of reactant and product compartment contents that define a \textit{distinguishable} instance of class $c$.
Two transitions within a class are called distinguishable if they are associated with a different stoichiometry when expressed in the form~\ref{eq:ArbitraryPopulationTransition}.
In our settings, this practically means that, whenever $\NumReact_c$ or $\NumProd_c$ take value $2$, we will count only pairs of $\xc_c$ or $\yc_c$ that are combinatorially distinct, because the ordering of the compartments is physically irrelevant. 
In order to formally enumerate the distinct transitions within a class, we introduce a bijective mapping $j=\varphi_c(\xc_c,\yc_c)$ that assigns to each distinguishable choice of $\xc_c$ and $\yc_c$ a unique index $j$ (and vice versa). This index $j$ refers to a specific stoichiometric equation of the form~\ref{eq:ArbitraryPopulationTransition}, whose stoichiometric arrays $\ReactantSv^c_j$ and $\ProductSv^c_j$ take entries $\ReactantS^c_{j,\vb{x}} = \sum_{\vb{z}\in\xc_c} \delta_{\vb{x},\vb{z}}$ and $\ProductS^c_{j,\vb{x}} = \sum_{\vb{z}\in\yc_c} \delta_{\vb{x},\vb{z}}$, with the symbol $\delta$ denoting a Kronecker delta.
In practice, the mapping $\varphi_c$ can be made explicit by enumerating all the possible contents $\xc_c$ and $\yc_c$ without repeating indistinguishable instances. 
In the following, we will denote the image of $\varphi_c$ with $\Jset_c$, which collects all the transitions belonging to class $c$.

The second step in defining a transition class is the specification of a rate law that assigns a rate to each individual transition within the class as a function of its particular compartment contents $\xc_c$ and $\yc_c$. 
This rate law can be defined in two parts.
First, we introduce a probability per unit time for the reactant compartment(s) with content $\xc_c$ to participate in a transition of class $c$, given the current state $\vn$ of the population, i.e. 
\begin{align}
&P(\xc_c \text{ participating during } \d t \mid \vn) = \notag \\
&= k_c  g_c(\xc_c) w(\vn ; \ReactantSv_j^c) \d t
\label{eq:participation_rate}
\end{align}
with $k_c \in \mathbb{R}^+$ being a content-independent rate constant and $g_c(\xc_c)$ a positive-valued function which tunes the rate in terms of the reactant compartment content(s). Note that $g_c(\xc_c)$ must be symmetric in its arguments when $\NumReact_c=2$.
The term $w(\vn;\ReactantSv_j^c)$ is a population weight that takes into account all the possible ways the current state $\vn$ could realize a transition involving reactant compartments with content $\xc_c$. 
In this work we consider $w(\vn;\ReactantSv_j^c)$ to be a combinatorial weight that reflects the physical indistinguishability of compartments having equal content. Thus, we set 
\begin{equation}
    w(\vn ; \ReactantSv_j)  = \prod_{\vb{x}\in\SpaceX} \binom{\nx{\vb{x}}}{\ReactantS^c_{j,\vb{x}}} ,
\label{eq:MassAction}
\end{equation}
which is analogous to the mass-action principle of standard stochastic reaction kinetics. 
Note that the binomials in eq.~\ref{eq:MassAction} take value different from $1$ only when $\vb{x} \in \xc_c$.

Finally, the second part of the rate law is a conditional probability $\pi_c(\yc_c\mid\xc_c)$ that describes how likely the reacting compartments of content $\xc_c$ result in product compartments of content $\yc_c$.
While this step might be generally of probabilistic nature, it can also be used to encode deterministic outcomes. For instance, coming back to the example of compartment fusion, the content of the product compartment is uniquely determined by fixing the contents of the two reactant compartments. 
Similarly to $g_c(\xc_c)$, we require $\pi_c$ to be symmetric in $\xc_c$ or $\yc_c$ whenever either of those involves two compartments.
In summary, eq.~\ref{eq:participation_rate} and the outcome distribution $\pi_c$ determine the propensity function of a specific transition $j$ of class $c$, i.e.
\begin{equation}
   \prop_{c,j}(\vn ) =  k_c  g_c(\xc_c) w(\vn ; \ReactantSv_j^c) \pi_c(\yc_c\mid\xc_c) ,
   \label{eq:ClassPropensity}
\end{equation}
which involves compartments $\{\xc_c,\yc_c \}= \varphi_c^{-1}(j)$.
The rate law~\ref{eq:ClassPropensity} provides a versatile definition, which allows us to equip a transition class with different physical properties, constraints or selectivity. 
We emphasize that eq.~\ref{eq:ClassPropensity} can be parameterized entirely in terms of the involved contents $\xc_c$ and $\yc_c$, such as illustrated for the examples in Table~\ref{tab:TransitionExamples}.
Additional information on how chemical reactions can be described as compartment transition classes can be found in SI Appendix S.1.

We can now reformulate the general counting process model from eq.~\ref{eq:dn} using the concept of transition classes. In particular, we associate a counter $R_{c,j}(t)$ with each transition  $j \in \Jset_c$ within class $c$. Moreover, we introduce the total class-transition counter $R_c(t)= \sum_{j\in\Jset_c} R_{c,j}(t)$, which corresponds to the cumulative number of events associated with class $c$ that happened until time $t$. The rate function of $R_c(t)$ is given by the total propensity function $\Prop_c(\vn) = \sum_{j\in\Jset_c} \prop_{c,j}(\vn)$, which follows from the superposition theorem for Poisson processes~\cite{Andersen2012}. The propensity $\Prop_c(\vn)$ represents the probability per unit time of any event in class $c$ to occur, given the current state $\vn$. 
Based on this, we can rewrite~eq.~\ref{eq:dn} as
\begin{align}{\label{eq:dn_classes_full}}
	\d\vn &= \sum_{c \in \Cset} \sum_{j \in \Jset_c} \Delta\vn_j^c \mathrm{d}R_{c,j}, \\
	         &= \sum_{c \in \Cset}  \Delta\vn^c \mathrm{d}R_{c},
\label{eq:dn_classes}
\end{align}
where $\Cset$ is a finite set of transition classes.
Whenever a transition in class $c$ occurs ($\d R_c = 1$), the state $\vn$ changes by a random state update $\Delta\vn^c$ with distribution $P(\Delta\vn^c=\Delta\vn_j^c \mid \vn) = \prop_{c,j}(\vn) /\Prop_c(\vn)$.
The jump-process representation in eq.~\ref{eq:dn_classes} is analytically convenient, and moreover entails an efficient strategy to perform stochastic simulations of population dynamics (see SI Appendix S.2).

\subsection{Stochastic moment dynamics}
We next show how the counting process model from eq.~\ref{eq:dn_classes_full} can be used to characterize the dynamics of the summary statistics of the compartment population.
A moment associated with the population state $\vn$ can be defined as
\begin{equation}
    \Momx{\gamma} = \sum_{\vb{x} \in \SpaceX} \vb{x}^\gamma \nx{\vb{x}} ,
\label{eq:MomentsDefinition}
\end{equation}
where $\vb{x}^\gamma = \prod_{i=1}^D x_i^{\gamma_i}$, with $\gamma$ being a vector of non-negative integer exponents. The sum $\sum_i \gamma_i$ sets the order of the moment $\Momx{\gamma}$. For instance, if $\sum_i \gamma_i=0$ then the moment corresponds to the total number of compartments present in the population, i.e., $N=\Momx{\mathbf{0}}=\sum_{\vb{x}} \nx{\vb{x}}$. 
Similarly, moments of order $1$ represent the total amount -- or mass -- of a particular species and so forth. It is important to keep in mind that the compartment number distribution $n(\vb{x})$ is stochastic and therefore each population moment will be stochastic as well. In the following, our goal is to derive an equation which captures the stochastic moment dynamics.

We begin by studying how a single transition $j \in \Jset_c$ of transition class $c \in \Cset$ affects an arbitrary population moment. 
Assume that right before the transition the population is in configuration $\vn^-$, and consider an associated moment $\Momx{\gamma, -}$. 
When the transition happens, the moment instantaneously changes to
\begin{align}
\Momx{\gamma, +} &= \sum_{\vb{x} \in \SpaceX} \vb{x}^\gamma\left( n^-(\vb{x}) +  \Delta n_{j,\vb{x}}^c\right) =\Momx{\gamma, -} + \sum_{\vb{x} \in \SpaceX} \vb{x}^\gamma \Delta n_{j,\vb{x}}^c \notag \\
&= \Momx{\gamma, -} + \Delta\Momx{\gamma}_{c,j},
\end{align}
with $\Delta\Momx{\gamma}_{c,j}$ as the net change of moment $\Momx{\gamma}$ due to transition $j$ of class $c$.
Correspondingly, we can write the differential change of any population moment in point process notation
\begin{align}{\label{eq:dM}}
	\d\Momx{\gamma} &= \sum_{c \in \Cset} \sum_{j \in \Jset_c }  \Delta\Momx{\gamma}_{c,j } \d R_{c,j } \\
				     &= \sum_{c \in \Cset}  \Delta\Momx{\gamma}_c \d R_c,
\label{eq:ClassSDE}
\end{align} 
where, in the second line, $\Delta\Momx{\gamma}_c$ is a random jump update with distribution $P(\Delta\Momx{\gamma}_c = \Delta\Momx{\gamma}_{c, j} \mid \vn) = \prop_{c,j}(\vn)/\Prop_c(\vn)$, analogously to eq.~\ref{eq:dn_classes}.

We finally remark that a useful distinction between transition classes can be made based on the moment updates $\Delta N_{c} =\Delta\Momx{\mathbf{0}}_{c}=\NumProd_c - \NumReact_c$ related to the total compartment number $N$. In particular, a transition class corresponding to a chemical modification inside one or more compartments will leave the total number of compartments $N$ unaffected and thus, $\Delta N_{c}=0$. In the following, we will refer to such transition classes as \textit{chemical events}. All other cases are referred to as \textit{compartment events}, since for those transition classes $\Delta N_{c}\neq 0$.

\subsection{Calculating mean and variance of the population moments}

To effectively describe fluctuations in the population moments $M$, we derived ordinary differential equations that capture the time evolution of their average and variance. We show in SI Appendix S.3 that the expectation of an arbitrary population moment satisfies the equation
\begin{align}
    \frac{\d}{\d t }\Et{\Momx{\gamma}} &=   \sum_{c\in\Cset} \Etb{\sum_{j\in\Jset_c} \Delta\Momx{\gamma}_{c, j} \prop_{c,j}(\vn) },
\label{eq:ExpectedMoments}
\end{align}
where $\langle \cdot \rangle$ denotes the expectation operator. 
Note that, since the moment change $\Delta\Momx{\gamma}_{c, j}$ is a constant for each $j \in \Jset_c$, the expectation could in principle be moved inside the second sum in~\ref{eq:ExpectedMoments}. 
In the latter form, however, the r.h.s. of~\ref{eq:ExpectedMoments} involves infinite sums of moments of the number distribution $n(\vb{x})$ itself, which defeats the purpose of a low-dimensional description in terms of population moments. 
Instead, we show that under certain conditions the sum $\sum_{j\in\Jset_c} \Delta\Momx{\gamma}_{c, j} \prop_{c,j}(\vn)$ can be expressed again as a function of a finite number of population moments, so that, after applying the expectation operator, eq.~\ref{eq:ExpectedMoments} reduces to a self-contained system of differential equations. A sufficient condition for this to be the case is that i) the function $g_c$ is a polynomial in $\xc_c$ and ii) the conditional distribution $\pi_c$ has moments which are polynomials in $\xc_c$ too (see SI Appendix S.4). In the present study, we will focus on systems which exhibit those two properties.

Analogously to eq.~\ref{eq:ExpectedMoments} for the expected moment dynamics, we can derive differential equations for the expectation of a squared moment using the rules of stochastic calculus for counting processes (see SI Appendix S.5). In combination with~\ref{eq:ExpectedMoments}, this  allows us to study the expected behavior of a population moment as well as its variability across different realizations. 
This is an important difference from conventional mean-field approaches, in which fluctuations in the number distribution and their corresponding moments are neglected. 

We finally remark that the coupled ODE system resulting from eq.~\ref{eq:ExpectedMoments} will not be closed in general since its r.h.s. may depend on higher-order moments. 
This problem can be addressed using moment-closure approximations, where moments above a certain order are approximated by functions of moments up to that order \cite{Singh2010}. These approximation schemes typically rely on certain assumptions on the underlying process distribution and may give more or less accurate results depending on the details of the considered system \cite{Schnoerr2014}. 
In our analyses, we found the multivariate Gamma closure as proposed in~\cite{Lakatos2015_MultivariateClosure} to give accurate results and we will adopt this choice of closure in our case studies when needed (details in SI Appendix S.6).

\section{CASE STUDIES}
We next demonstrate our framework and the moment equation approach using several case studies inspired by biological systems at different scales.
All simulations have been performed using the scientific computing language julia \cite{julia}. 

\subsection{Nested birth-death process}
\begin{figure*}[t]
\centering
\includegraphics[width=.95\linewidth]{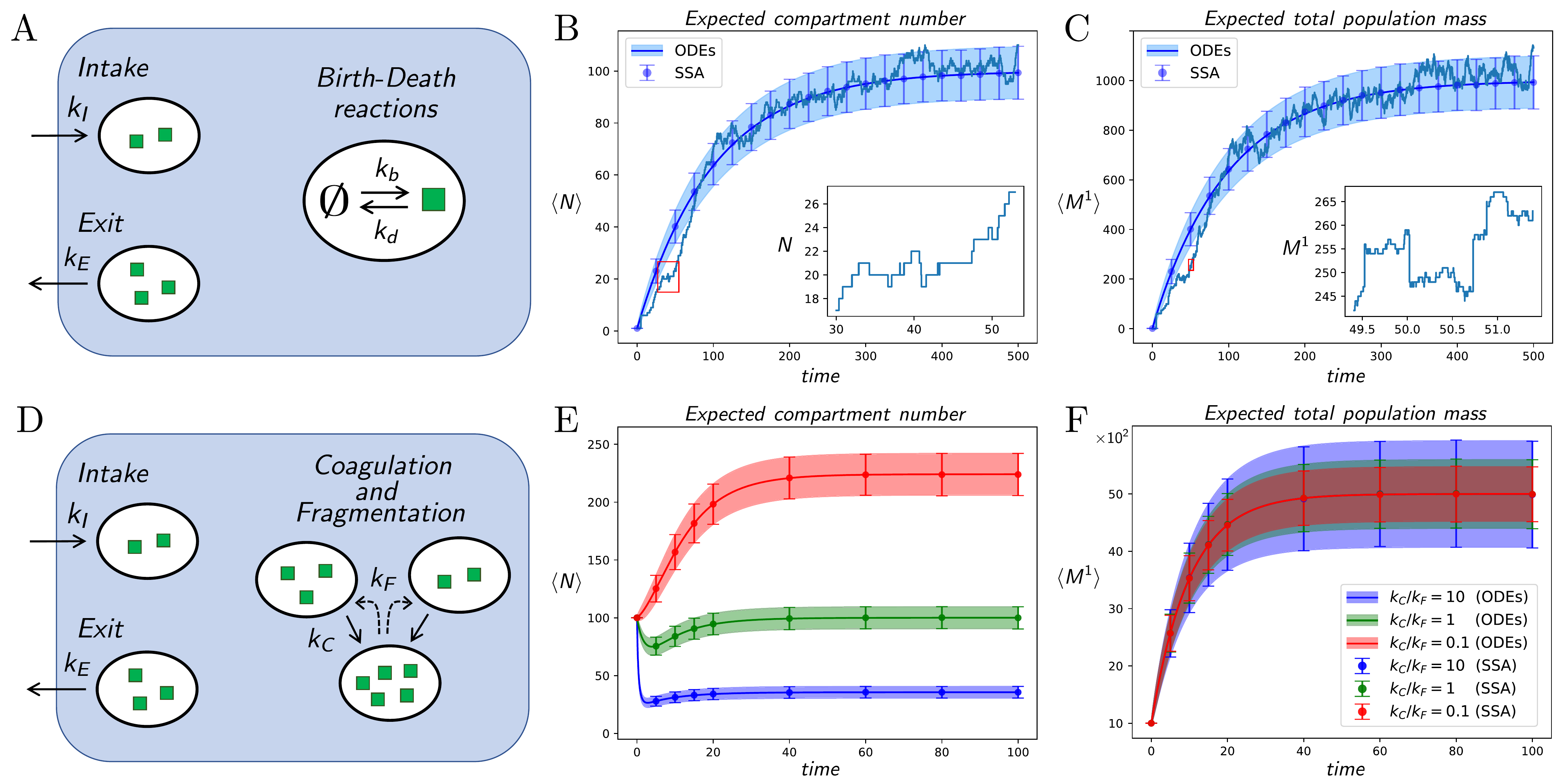}
\caption{A) Schematic illustration of the nested birth-death system.
B-C) Expected dynamics of compartment number and total mass, obtained from moment equations (ODEs) and stochastic simulations (SSA). Errorbars and shaded areas correspond to one standard deviation above and below the mean. The superimposed blue line shows one stochastic realization, of which a small section (highlighted in red) is enlarged in the insets to illustrate the stochastic jump dynamics. 
D) Schematic illustration of the coagulation-fragmentation model.
E-F) Expected dynamics of compartment number and total population mass for different coagulation rates.}
\label{fig:BD_CF}
\end{figure*}
We begin by considering a population of compartments with univariate content $x\in[0,\infty)$ and introducing a simplistic toy model defined by the four transition classes
\begin{align}
    \emptyset &\xrightharpoonup{h_I(\vn;y)} [y] && h_I(\vn;y)=k_I\pi_{Poiss}(y;\lambda) \notag \\
    [x]   &\xrightharpoonup{h_E(\vn;x)} \emptyset && h_E(\vn;x)=k_E\nx{x} \notag \\
    [x] &\xrightharpoonup{h_b(\vn;x)} [x+1] && h_b(\vn;x)=k_b \nx{x}  \notag \\
    [x] &\xrightharpoonup{h_d(\vn;x)} [x-1] && h_d(\vn;x)=k_d x \nx{x},
    \label{SCN:BirthDeath}
\end{align}
which are also illustrated in Fig.~\ref{fig:BD_CF}A.
The first two transition classes in~\ref{SCN:BirthDeath} are respectively an intake transition class, where a new compartment enters the population with a Poisson-distributed content with mean-parameter $\lambda$, and a random-exit transition class, for which any compartment can leave the population with the content-independent exit rate $k_E$. 
According to our terminology, these first two transitions classes are compartment events, since they affect the number of compartments in the population, whereas the last two transition classes in~\ref{SCN:BirthDeath} account for chemical modifications. 
The total propensities associated with~\ref{SCN:BirthDeath} are found to be $\Prop_I(\vn)=k_I$, $\Prop_E(\vn)=k_E N$, $\Prop_b(\vn)=k_b N$ and $\Prop_d(\vn)=k_d \Momx{1}$. 
We start writing the stochastic differential equation for the number of compartments $N$ in the form~\ref{eq:ClassSDE} 
\begin{equation}
    \d N = \d R_I - \d R_E, \\
    \label{eq:SDE_N_BirthDeath}
\end{equation}
which is affected only by the occurrence of compartment events, while the chemical birth-death events do not alter $N$. 
For the total population mass $\Momx{1}=\sum_{x=0}^\infty x\nx{x}$, we can use~\ref{eq:dM} to find
\begin{align}
\d\Momx{1} =& \sum_{y=0}^\infty (+y)\d R_{I,y} + \sum_{x=0}^\infty (-x) \d R_{E,x} \notag \\
&+ \sum_{x=0}^\infty (+1) \d R_{b,x} +\sum_{x=0}^\infty (-1) \d R_{d,x} .
 \label{eq:SDE_M_BirthDeath}
 \end{align}
Note that the mass updates related to intake or exit events depend on the content of each specific transition, while for birth or death events they always take values $+1$ or $-1$, respectively.
We can express eq.~\ref{eq:SDE_M_BirthDeath} in the compact form~\ref{eq:ClassSDE} too, which equals
\begin{equation}
    \d \Momx{1} = Y_I\d R_I -X_E\d R_E + \d R_b - \d R_d, 
    \label{eq:SDE_M_BirthDeath_compact}
\end{equation}
where we introduced the random variables $\Delta\Momx{1}_I=Y_I \sim \pi_{Poiss}(y;\lambda)$ and $\Delta\Momx{1}_E=-X_E$ with $\P(X_E=x \mid \vn)=\nx{x}/N$, which is a categorical distribution for the content of the compartment randomly exiting the system, found through $h_E(\vn;x)/H_E(\vn)$. 
We can proceed to study the average trajectory for eqs.~\ref{eq:SDE_N_BirthDeath} and~\ref{eq:SDE_M_BirthDeath} by using the result~\ref{eq:ExpectedMoments}. 
We obtain
\begin{align}
\frac{\d \Et{N}}{\d t} &= k_I - k_E\Et{N} \notag \\
\frac{\d \Et{\Momx{1}}}{\d t} &= k_I\lambda - k_E\Et{\Momx{1}} + k_b\Et{N} - k_d\Et{\Momx{1}} .
\label{eq:ODE_N_M_BirthDeath}
\end{align}{}
Not surprisingly, the evolution of $\Et{N}$ is independent of that of $\Et{\Momx{1}}$, as can be seen already by inspection of the total propensities. 
Indeed, in this example the dynamics of $N$ is a birth-death process with rates $k_I$ and $k_E$. 
On the contrary, the expected total mass $\Et{\Momx{1}}$ evolves in a coupled fashion with the number of compartments. 
We further derive equations for the variability of $N$ and $\Momx{1}$ around their average trends (SI Appendix S.7.1). 
In summary, this leads to a system of $6$ coupled ODEs which can be integrated numerically to compute the exact trajectories for the expected mean and variability of $N$ and $\Momx{1}$, as shown in Fig.~\ref{fig:BD_CF}B-C. 
Note that no moment-closure approximation is required for this system.

From the moment equations we can further derive some analytical results regarding the steady-state properties of the compartment population.
As expected, the steady-state number of compartments exhibits Poisson statistics with mean and variance $k_I/k_E$ (SI Appendix S.7.2).
Denoting for compactness $\Et{\Momx{1}_\infty}= \lim_{t\rightarrow\infty}\Et{\Momx{1}}$, the steady-state expected total mass equals
\begin{equation}
    \Et{\Momx{1}_\infty}=\frac{k_I}{k_E}\bqty{\frac{k_b}{k_d}\frac{1+\alpha\beta}{1+\alpha}} ,
    \label{eq:BD_mass_ss}
\end{equation}
where the term in square brackets corresponds to the average steady-state \textit{per-compartment} content $\Et{X_\infty}$, and we introduced the dimensionless parameters $\alpha=k_E/k_d$ and $\beta=\lambda/(k_b/k_d)$. 
Note that, for all $\alpha>0$, setting $\beta=1$ in eq.~\ref{eq:BD_mass_ss} gives $\Et{X_\infty}=k_b/k_d$, which is consistent with the steady-state of the chemical birth-death process occurring in each compartment. In other words, for $\lambda=k_b/k_d$ the content of new compartments entering the system exhibits the same Poisson distribution of a single birth-death process with rates $k_b $ and $k_d$ at stationarity, thereby preserving Poissonian statistics across the compartment contents. 
This result is resembled also by the analytical expression for the variance-to-mean ratio of $X_\infty$, which exhibits a global minimum at value $1$ for $\beta=1$ (see SI Appendix S.7.2. and Fig.~S.1). 
This simple case study serves to illustrate how the proposed framework can be used to study fluctuations in systems that exhibit both compartment and reaction dynamics. In the following we will consider systems with more complex interactions. 

\subsection{Stochastic coagulation-fragmentation dynamics}
Coagulation-fragmentation (CF) processes form an important class of models to describe populations of interacting components \cite{Krapivsky2010}. 
CF models have been used to study biological phenomena at different scales, including  protein clustering~\cite{Saunders2015}, vesicle trafficking \cite{Foret2012,Vagne2018,Vagne2018sorting} or clone-size dynamics during development~\cite{Rulands2018}. 
Previously, these models have been analyzed mostly using mean-field approaches or forward stochastic simulation.  
In this case study, we will revisit these models by analyzing fluctuations in their dynamics using the proposed moment-equation approach.
For simplicity, we will consider again a univariate compartment content $x\in[0,\ldots,\infty)$, but we remark that multivariate scenarios can be handled analogously.
We define a random coagulation class, where each pair of compartments is equally likely to fuse with rate $k_C$ (i.e., $g_C(x,x')=1$ in Table~\ref{tab:TransitionExamples}).
Instead, we introduce for instance a mass-driven fragmentation class, where a compartment undergoes a fragmentation event with rate $k_Fg_F(x)=k_Fx$ that is proportional to its content. For $\pi_F(y \mid x)$ we choose a uniform fragment distribution.
The corresponding total class propensities read $\Prop_C(\vn)=k_CN(N-1)/2$ and $\Prop_F(\vn)=k_F\Momx{1}$  (see SI Appendix S.8.1). 
Moreover, we might consider that the population can exchange compartments with an external environment. 
In order to account for this, we can equip our model with an intake and an exit transition class, similarly to model~\ref{SCN:BirthDeath}. 
Considering these four transition classes (Fig.~\ref{fig:BD_CF}D), the SDE for $N$ is
\begin{equation}
    \d N = \d R_I - \d R_E - \d R_C + \d R_F,
    \label{eq:CF_N_SDE}
\end{equation}
because $\Delta N$ is equal to $-1$ for any exit or coagulation event and $+1$ for intake or fragmentation. 
The SDE for the total mass $\Momx{1}$ of the compartment population assumes an even simpler form
\begin{equation}
    \d \Momx{1}= Y_I\d R_I - X_E\d R_E ,
    \label{eq:CF_M_SDE}
\end{equation}
since coagulation and fragmentation events conserve mass. 
The random variables $Y_I$ and $X_E$ are defined like in the previous case study. 
For space considerations, the derivation of the moment equations is left to SI Appendix Section S.8.1. 
In particular, since the considered system does not exhibit closed moment-dynamics, we made use of the proposed Gamma closure as mentioned earlier. 
In Fig.~\ref{fig:BD_CF}E-F we plot the expected trajectories and fluctuations of $N$ and $\Momx{1}$ for different parameter settings and compare them to exact stochastic simulations. In all cases, we found very good agreement between both approaches. An analogous analysis of the expected second order moment is provided in SI Appendix Fig.~S.2.
An interesting feature emerging from Fig.~\ref{fig:BD_CF}F is that the coagulation- and fragmentation rates affect the variability of the total mass, but not its average behavior, which depends only on the intake and exit parameters. 
This happens because a larger coagulation rate implies that the same total mass has to be shared among fewer compartments, which causes the total population mass to exhibit larger fluctuations upon occurrence of the intake and exit events. 
This fact could not be captured by a mean-field treatment of a coagulation-fragmentation system, since fluctuations are necessarily lost in that case. 
Similar considerations hold true for the expected compartment number dynamics
\begin{equation}
    \frac{\d \Et{N}}{\d t} = k_I - k_E\Et{N} -\frac{k_C}{2}\pqty{\Et{N^2}-\Et{N}}+k_F\Et{\Momx{1}} ,
    \label{eq:CF_N_ODE}
\end{equation}
where we point out the dependency on the second-order moment $\Et{N^2}$. 
In a mean-field approximation, the coagulation term in eq.~\ref{eq:CF_N_ODE} would simplify to $-\frac{k_C}{2}\Et{N}^2$.
Indeed, replacing $\Et{N^2}$ with $\Et{N}^2$ implies that $\mathrm{Var}(N)=\Et{N^2}-\Et{N}^2=0$, thereby neglecting fluctuations in the compartment number. 
Additionally, the linear correction $k_C\Et{N}/2$ in eq.~\ref{eq:CF_N_ODE}, which originates from the exact combinatorics of the possible compartment pairings, would be omitted too. 
Both of these approximations can lead to significant deviations when only few compartments are present in the system. 
For more details on the validity of mean-field approximations in stochastic coagulation systems, the reader may refer to~\cite{Tanaka1993}. 

\subsection{Transcription dynamics in a cell community}
\label{sec:cell_communication}
\begin{figure*}[t]
\centering
\includegraphics[width=.95\linewidth]{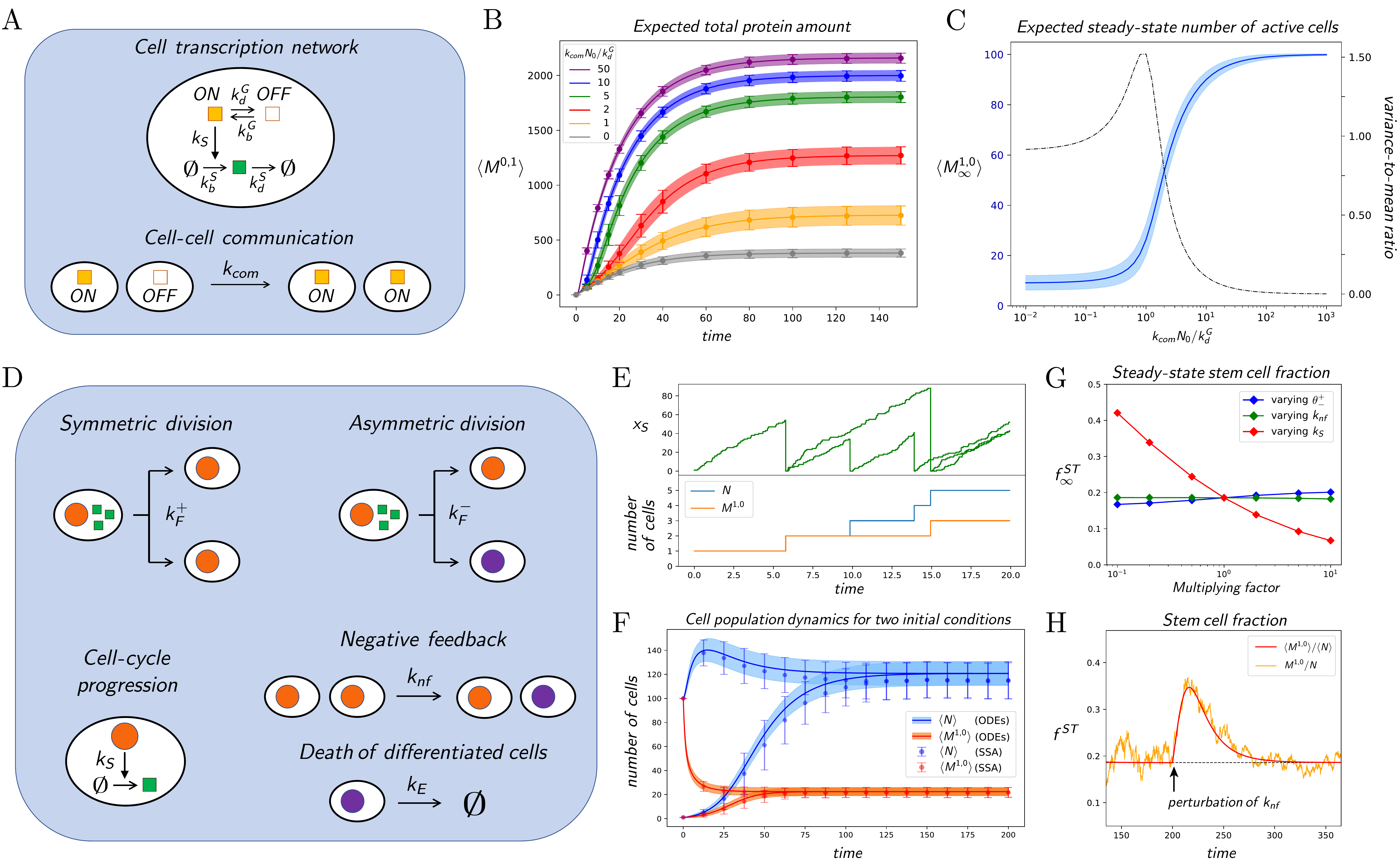}
\caption{A) Schematic illustration of the cell communication model.
B) Expected dynamics of the total protein mass $\Momx{0,1}$ for different values of $k_{com}$, with one standard deviation above and below the mean.
Lines and shaded areas correspond to the result of moment equations, while dots and error bars were obtained from $10^3$ stochastic simulations. 
At time zero, only one cell is in the active state.
C) Expected steady-state number of active cells and corresponding variance-to-mean ratio $\mathrm{Var}(\Momx{1,0}_\infty)/\Et{\Momx{1,0}_\infty}$ for different values of $k_{com}$, plotted as a function of $k_{com}N_0/k_d^G$.
D) Schematic illustration of the stem cell model. 
E) The accumulation-reset stochastic dynamics of $x_S$ in stem cells is shown for the initial transient of a single realization, starting with one stem cell. The lower panel shows the corresponding changes in total cell number and stem cell number.
F) Comparison of the expected dynamics for the total cell number $N$ (blue) and stem cell number $\Momx{1,0}$ (orange) from moment equations (ODEs) and stochastic simulations (SSA), for two different initial conditions.
G) Dependency of the steady-state stem cell fraction on variations of some model parameters, computed from moment equations. 
H) Robust dynamics of the stem cell fraction, upon applying a perturbation at time $t=200$  where $k_{nf}$ was suddenly reduced by a factor $5$. 
In red, the expected stem cell fraction obtained from moment equations. In orange, one particular stochastic realization.
}
\label{fig:cell_communication}
\end{figure*}

In our next case study we apply our approach to analyze a population of compartments which are chemically coupled to each other. 
To this end, we consider a cell population of fixed size $N_0=100$, and we equip each cell (i.e. each compartment) with a transcription network, as shown in Fig~\ref{fig:cell_communication}A. 
A binary gene variable can stochastically switch between an off state $(x_G=0)$ and on state $(x_G=1)$. 
The active state promotes the production of a protein $S$ at rate $k_S$. 
Furthermore, the species $S$ is constantly produced at a low leakage rate $k_b^S \ll k_S $ and degrades at rate $k_d^S$. 
We define the two-dimensional compartment content variable $\vb{x}=(x_G,x_S) \in \SpaceX = [0,1] \times \Nat$. 
To account for cell-to-cell communication, we consider a messenger molecule which is released by cells in the active state and promotes the activation of the gene in inactive cells.
In the limit of fast diffusion, we can effectively describe this communication mechanism as a bi-compartmental transition class
\begin{equation}
	[\vb{x}] + [\vb{x'}] \xrightharpoonup{\prop_{com}(\vn;\vb{x},\vb{x'})} [1,x_S] + [1,x_S'] ,
	\label{eq:CommEvent}
\end{equation}
where we made explicit in the r.h.s. that, upon this transition, both cells are in the active state (details in SI Appendix S.9).
The total class propensity for~\ref{eq:CommEvent} equals
\begin{equation}
	\Prop_{com}(\vn) = k_{com} \Momx{1,0}\pqty{N_0-\Momx{1,0}},
	\label{eq:CommTotProp}
\end{equation}
which reflects the fact that the global activation rate is proportional to the product of the number of active cells $\Momx{1,0}$ and the number of inactive cells (i.e. $N_0-\Momx{1,0}$) in the current configuration $\vn$. 
Note that in the considered model all transition classes are chemical transitions according to our definition, so that the number of compartments remains constant at its initial value $N_0$. 
We are now interested in studying the transcription activation dynamics in the cell population, as a function of the communication rate constant $k_{com}$. 
To this end, we derived moment equations describing the averages and variances of the active cell number $\Momx{1,0}$ and the total amount of transcribed proteins $\Momx{0,1}$ (SI Appendix S.9.1).  

In Fig.~\ref{fig:cell_communication}B we plot the expected total protein dynamics $\Et{\Momx{0,1}}$ for different values of the communication rate $k_{com}$. 
As before, results obtained from exact SSA and the moment-based approach are in very good agreement with each other.  
The expected dynamics of the number of active cells $\Et{\Momx{1,0}}$ is shown in SI Appendix Fig.~S.3. 
Moreover, we show the steady-state expected number of active cells $\Et{\Momx{1,0}_\infty}$ as well as its variance-to-mean ratio as a function of $k_{com}$ in Fig~\ref{fig:cell_communication}C. 
Interestingly, the noise in the steady-state number of active cells initially increases with $k_{com}$, peaks around $k_{com} N_0 / k_d^G =1$ and then soon starts dropping towards zero as the activation saturates. 
As can be seen from this analysis, moment equations provide an effective means to access the statistics of a compartment population for a wide range of parameters and identify interesting dynamical regimes with little computational effort.

\subsection{Stem cell population dynamics}
In our last application we aim to study a system involving a more complex interplay of chemical and compartmental dynamics. 
In particular, we consider a model inspired by the proliferation and differentiation dynamics of stem cell populations \cite{Stiehl2011,Yang2015}. 
In our model (Fig.~\ref{fig:cell_communication}D) each compartment represents one cell, whose content is described by $\vb{x}=(x_G,x_S) \in \SpaceX = [0,1] \times \Nat$, as in the previous case study.
Here, the binary content-variable $x_G$ indicates if a cell is either a stem cell ($x_G=1$) or a differentiated cell ($x_G=0$), while $x_S$ represents the abundance of a factor that controls the rate of cell division. 
In each stem cell, $x_S$ stochastically increases over time at rate $k_S$ and is reset to zero in both daughter cells after a cell division occurs. Therefore, $x_S$ can be interpreted as a proxy for cell-cycle stage. This accumulation-reset mechanism is illustrated in Fig.~\ref{fig:cell_communication}E, where the dynamics of $x_S$ across a lineage is followed over multiple rounds of cell division along a stochastic realization. 

In our model, each cell division can lead to two random outcomes: either a symmetric division, where both daughter cells are stem cells, or an asymmetric division, where one of the daughter cells differentiates. Additionally, we introduce a negative feedback mechanism, which causes stem cells to differentiate at a rate that increases with their own abundance. 
Phenomenologically, we can account for negative feedback by introducing a second-order compartment event, which mimics the interaction of a stem cell with the remaining stem-cell population. 
Finally, we assume that differentiated cells die or exit the system at a constant rate $k_E$. 

Our goal is to study the dynamics and the variability of the total cell number $N$ and the stem cell number $\Momx{1,0}$ in the population.
We remark that, in comparison to the previous case studies, the application of the moment equation method turns out to be more challenging for this model. 
This is because the total propensities of the division events depend on the second-order moment $\Momx{1,1}=\sum_{\vb{x}} x_G x_S$, which represents the total amount of $x_S$ in stem cells. 
In combination with the second-order feedback mechanism, this would lead to a large number of equations required to capture the full dynamics of all involved moments up to a certain order. Here we address this problem by combining the multivariate Gamma closure with a mean-field approximation, where correlations among certain population moments are neglected (SI Appendix S.10.1).

In Fig.~\ref{fig:cell_communication}F we plot the expected number dynamics starting from two different initial conditions ($1$ or $100$ stem cells) and compare it to stochastic simulations. 
Even though we used additional approximations, the moment dynamics are in relatively good agreement with the results obtained from stochastic simulations.
Based on the moment-equations, we next investigated how the steady-state stem cell fraction $f^{ST}_\infty=\Et{\Momx{1,0}_\infty}/\Et{N_\infty}$ is affected by varying three different parameters of the model: the feedback rate $k_{nf}$, the rate $k_S$ and the ratio $\theta^+_-=k_F^+/k_F^-$, with $k_F^+ + k_F^-$ held constant (Fig.~\ref{fig:cell_communication}G). 
Interestingly, we find that the stem-cell fraction is largely robust against changes in the feedback strength $k_{nf}$ as well as the ratio of division rates $\theta^+_-$.  
In regard of the former, while changing $k_{nf}$ affects the number of stem cells present in the system (SI Appendix Fig.~S.4), the relative speed between symmetric and asymmetric divisions remains unaffected, thereby preserving the total stem cell fraction.
This is further illustrated in Fig.~\ref{fig:cell_communication}H which shows how the stem-cell fraction returns to its set point upon perturbing $k_{nf}$. 
Instead, considering variations of $\theta^+_-$, the robustness of $f^{ST}_\infty$ seems to originate from the fact that the rate of symmetric divisions and feedback events compensate for each other. A more detailed analysis of the principles underlying the robustness properties of such models shall be performed in future works.
This last application shows that, even though approximate, the moment equation approach provides valuable insights into the collective dynamics of cell populations. 

\section{Discussion}
Compartmentalization of biochemical processes is a hallmark of living systems across different scales, from organelle dynamics to cell populations.
Theoretical approaches which address the interplay of compartment dynamics and molecular noise are therefore of great relevance.
In this work we introduced a general mathematical framework to model arbitrary compartmental and biochemical dynamics in a population of interacting compartments.
Our approach relies on a fully stochastic treatment and is thus suitable to investigate the effect of mesoscopic fluctuations on compartmentalized biochemical systems. 
We showed how the dynamics of a compartment population can be compactly described by ordinary differential equations, which capture means and variances of certain population moments, such as the compartment number or total molecular content. 
Therefore, this technique provides an analytical and computational means to efficiently access the statistical properties of the population, which could be otherwise obtained only through forward stochastic simulation.
One limitation of our approach is that it relies on the availability of suitable moment-closure approximations. In all our case studies, we found the Gamma-closure to give accurate results, but different closures may be required for other types of systems. 
In some of the presented case studies, we have shown how our framework can be used to track additional compartment properties in addition to their molecular content. 
For instance, compartments can be associated with distinct types or categories, each exhibiting different dynamical features.
This could be particularly relevant for studying stochasticity in developmental systems, where cells sharing the same progenity can commit to different fates and genetic programs.

\addtolength{\textheight}{-3cm}   


\section{Acknowledgments}
We thank Quentin Vagne for his helpful comments on the implementation of stochastic simulations.  
The authors were supported by core funding of the Max Planck Institute of Molecular Cell Biology and Genetics.


\end{document}


\maketitle

\section{Modeling chemical reactions as transition classes}
This section elaborates on the modeling of single-compartment chemical events as transition classes. 
Namely, we consider a stoichiometric equation of the form
\begin{equation}
\alpha_{l,1}X_1+\ldots+\alpha_{l,D}X_D \xrightharpoonup{\tilde{\prop}_l(\vb{x})} \beta_{l,1}X_1+\ldots+\beta_{l,D}X_D ,
\label{eq:StoichiometricEq}
\end{equation}
which can act on the content of each compartment in the population. 
We assume the propensity function $\tilde{\prop}_l$ to obey mass-action kinetics
\begin{equation}
\tilde{\prop}_l(\vb{x}) = k_l \prod_{i=1}^D \binom{x_i}{\alpha_{l,i}} = k_l g_l(\vb{x})
\label{eq:MassActionChem}
\end{equation}
with $k_l>0$ being the rate constant for the reaction~\ref{eq:StoichiometricEq}, which is indexed by $l$. 
Whenever the reaction $l$ occurs in a compartment of the population, its content $\vb{x}$ is updated by the change-vector $\Delta\vb{x}_l=\beta_{l} - \alpha_{l} $.
From the compartment population perspective, the occurrence of one chemical reaction $l$ in a single compartment of content $\vb{x}$ can be understood as an update of the population state $\vn$ where $\nx{\vb{x}}$ is decreased by $1$ and $\nx{\vb{x}+\Delta\vb{x}_l}$ is increased by $1$. 
The propensity function of such transition is given by the reaction propensity $\tilde{h}_l(\vb{x})$ times the factor $w(\vn)=\nx{\vb{x}}$,  which is the number of compartments with content $\vb{x}$ in which the reaction could occur.
Therefore, the stoichiometric equation~\ref{eq:StoichiometricEq} corresponds to the transition class
\begin{align}
[\vb{x}] &\xrightharpoonup{\prop_l(\vn;\vb{x})} [\vb{y}] , && \prop_l(\vn;\vb{x},\vb{y})=k_l g_l(\vb{x})\nx{\vb{x}} \delta_{\vb{y},\vb{x} + \Delta\vb{x}_l},
\label{eq:ChemicalTransitionClass}
\end{align}
where the product-compartment distribution is a Kronecker delta that accounts for a change in $\vb{x}$ by the state-change vector $\Delta\vb{x}_l$. This transition class can be equivalently expressed in a more compact form
\begin{align}
[\vb{x}] &\xrightharpoonup{\prop_l(\vn;\vb{x})} [\vb{x} + \Delta\vb{x}_l] , && \prop_l(\vn;\vb{x})=k_l g_l(\vb{x})\nx{\vb{x}} .
\end{align}
We remark that the rate constant and the content dependent function $g_l(\vb{x})$ of the transition class~\ref{eq:ChemicalTransitionClass} simply correspond to the single-compartment mass-action propensity defined in eq.~\ref{eq:MassActionChem}.
Furthermore, it is worth to emphasize that~\ref{eq:ChemicalTransitionClass} conserves the number of compartments in the population, consistently with our definition of chemical events. 
In summary, this shows how conventional stoichiometric equations of the form~\ref{eq:StoichiometricEq} can be included in a compartment population model, by expressing them in terms of transition class~\ref{eq:ChemicalTransitionClass}.

\section{Implementation of stochastic simulations}
In this section we explain how we implemented efficient stochastic simulations for compartment population models, which we used to evaluate the accuracy of the moment-based approach.
We recall that the next-event waiting time distribution and next-event class distribution are determined by the value of the total class propensities $\Prop_c(\vn)$ associated with the current state $\vn$.
For all classes $c$ admitting self-contained moment dynamics, the total class propensities are functions of some population moments.
We denote with $\mathcal{M}$ the set of those population moments that enter in the total class propensities $H_c(\vn)$, $c\in\Cset$, with $\Cset$ being the set of transition classes defining the population dynamics.  

A stochastic simulation with initial condition $\vn_0$ until time $t_{max}$ proceeds as follows:
\begin{enumerate}
\item Set $t \leftarrow 0$ and $\vn\leftarrow\vn_0$ 
\item Compute the moments $\Momset(\vn)$ 
\item Compute $\Prop_c\leftarrow\Prop_c(\Momset)$, $\forall c \in \Cset$, and $\Prop_{TOT}\leftarrow\sum_c \Prop_c$
\item Draw the next-event time as $t \leftarrow t + \pqty{-\frac{1}{\Prop_{TOT}}\log(1-U)}$, with $U \sim \text{Uniform}[0,1)$
\item If $t<t_{max}$,  \\ draw the next-event class $c^*$ from the discrete distribution $\text{Prob.}(c^*=c) = \frac{\Prop_c}{\Prop_{TOT}}$ \\
else,  \\  end the simulation.
\item Draw the reactant compartments so that their contents $\xc_{c^*}$ are sampled according to the distribution $\prop_{c^*}(\vn;\xc_{c^*})/\Prop_{c^*}(\vn)$ 
\item When needed, draw the product compartments $\yc_{c^*}$ from $\pi_{c^*}(\yc_{c^*} \mid \xc_{c^*})$
\item Update $\vn$ and $\Momset$ according to the drawn transition, and go to 3
\end{enumerate}

Note that the state of the system can be conveniently represented by storing the $D$-dimensional contents $\vb{x}_n$ of each compartment $n=1,\dots,N$ in a $D \times N$ integer matrix.

\section{Derivation of the expected moment dynamics}
Our goal is to derive the expected trajectory of the SDE 
\begin{align}
	\d\Momx{\gamma} &= \sum_{c \in \Cset} \sum_{j \in \Jset_c }  \Delta\Momx{\gamma}_{c,j } \d R_{c,j } ,
\label{eq:ClassSDE}
\end{align} 
which describes the time evolution of an arbitrary population moment $\Momx{\gamma}$ subject to a set $\Cset$ of transition classes.
Using the Doob-Meyer decomposition theorem~\cite{Andersen2012}, we can decompose the differential reaction counter $\d R_{c, j }$ into a predictable part, related to its propensity function $\prop_{c,j}(\vn)$, and a martingale $Q_{c, j }$
\begin{equation}
    \d R_{c, j } = \prop_{c,j}(\vn) \d t + \d Q_{c, j }.
    \label{eq:DoobMeyer}
\end{equation}{}
We can substitute the decomposition~\ref{eq:DoobMeyer} into~\ref{eq:ClassSDE} and take the expectation on both sides
\begin{align}
    \d\Et{\Momx{\gamma}} &= \Etb{\sum_{c\in\Cset} \sum_{j\in\Jset_c} \Delta\Momx{\gamma}_{c, j} \pqty{\prop_{c,j}(\vn) \d t + \d Q_{c, j }}} \notag \\
    &= \sum_{c\in\Cset} \sum_{j\in\Jset_c} \Delta\Momx{\gamma}_{c, j} \Et{\prop_{c,j}(\vn) \d t + \d Q_{c, j }} \notag \\
    &=   \sum_{c\in\Cset} \sum_{j\in\Jset_c} \Delta\Momx{\gamma}_{c, j} \Et{\prop_{c,j}(\vn)} \d t ,
\label{eq:ExpectedMoments_proof}
\end{align}
where the second term vanishes because of the martingale property of $Q_{c,j}$. 
Finally, we can rearrange the last line of~\ref{eq:ExpectedMoments_proof} to present it in the same form as the main paper
\begin{align}
    \frac{\d\Et{\Momx{\gamma}}}{\d t} &=  \sum_{c\in\Cset} \sum_{j\in\Jset_c} \Delta\Momx{\gamma}_{c, j} \Et{\prop_{c,j}(\vn)} \notag \\
    &=  \sum_{c\in\Cset} \Etb{\sum_{j\in\Jset_c} \Delta\Momx{\gamma}_{c, j} \prop_{c,j}(\vn)} .
\label{eq:ExpectedMoments}
\end{align}

\section{Conditions for self-contained moment dynamics}
In this section we provide sufficient conditions for a transition class to yield moment dynamics whose r.h.s. depends exclusively on population moments.
For an arbitrary population moment $\Momx{\gamma}$, the contribution of class $c$ to its expected dynamics is given by
\begin{align}
\frac{\d \Et{\Momx{\gamma}}}{\d t} &= \Etb{\sum_{j\in\Jset_c} \Delta\Momx{\gamma}_{c,j} \prop_{c,j}(\vn)} \notag \\
 &= \Etb{\sum_{\xc_c}\sum_{\yc_c} \Delta\Momx{\gamma}_{c}(\xc_c,\yc_c) \prop_{c}(\vn;\xc_c,\yc_c)} ,
 \label{eq:expected_xcyc}
\end{align}
where we have made explicit the reactant- and product- compartment contents $\xc_c$ and $\yc_c$ involved in each specific instance $j=\varphi_c(\xc_c,\yc_c)$ and accordingly parametrized the moment update and the propensity function. The sums over $\xc_c$ and $\yc_c$ are thus intended to enumerate all distinguishable transitions of class $c$.
If now we substitute the full expression of $\prop_c$ into~\ref{eq:expected_xcyc} we obtain
\begin{align}
\frac{\d \Et{\Momx{\gamma}}}{\d t} &= k_c \Etb{\sum_{\xc_c} w(\vn;\xc_c)g_c(\xc_c) \sum_{\yc_c} \Delta\Momx{\gamma}_{c}(\xc_c,\yc_c)\pi_c(\yc_c\mid\xc_c)} \notag \\
 &= k_c \Etb{\sum_{\xc_c} w(\vn;\xc_c)  g_c(\xc_c) \Et{\Delta\Momx{\gamma}_{c}(\xc_c,\yc_c) \mid \xc_c} },
 \label{eq:expected_full}
\end{align}
with $\Et{\cdot \,|\, \xc_c}$ as the conditional expectation with respect to $\pi_c$. 
We recall that $\Delta\Momx{\gamma}_{c}(\xc_c,\yc_c)$ is a polynomial in $\xc_c$ and $\yc_c$ by construction, because it is defined as a difference of discrete compartment contents raised to exponents $\gamma$.
Thus, the expectation $\Et{\Delta\Momx{\gamma}_{c}(\xc_c,\yc_c)\mid\xc_c}$ will in general contain terms of the form $\Et{\yc_c^\xi \mid \xc_c}$ for some $\xi_i\leq\gamma_i$, $i=1,\ldots,D$, that is, the conditional moments of the distribution $\pi_c$ up to order $\gamma$. 
Besides that, the combinatorial weight $w(\vn;\xc_c)$ has been defined as a product of binomials, which implies that it is a polynomial in $\nx{\vb{x}}$, for $\vb{x} \in \xc_c$.

At this point we can see under which condition the sum over $\xc_c$ will produce a function of some population moments on the right-hand side of~\ref{eq:expected_full}.
What we need to require is that the product $g_c(\xc_c) \Et{\Delta\Momx{\gamma}_{c}(\xc_c,\yc_c) \mid \xc_c}$ is a polynomial in $\xc_c$: in that case, the sum over $\xc_c$ next to the terms $\nx{\vb{x}}$ will result in a function of population moments. 
Thus, $g_c(\xc_c) \Et{\Delta\Momx{\gamma}_{c}(\xc_c,\yc_c) \mid \xc_c}$ being a polynomial is a sufficient condition for the propensity of class $c$ to admit a self-contained form.
For instance, this condition is satisfied if 
\begin{itemize}
\item $g_c(\xc_c)$ has polynomial dependency on $\xc_c$, and
\item the conditional moments of $\pi_c(\yc_c\mid\xc_c)$ are polynomials in $\xc_c$ ,
\end{itemize}
since a product of polynomials is in turn a polynomial.
This corresponds to the sufficient condition reported in the main text.

\section{Ito's rule for counting processes}
The result of Eq.~12 in the main text can be equivalently written also for functions of one or more population moments by means of the Ito's rule for counting processes~\cite{Oksendal2007}. 
Considering a function $f(\Momx{\gamma})$, its SDE is obtained as
\begin{equation}
    \d f(\Momx{\gamma}) = \sum_{c \in \Cset} \sum_{j \in \Jset_c} \bqty{f\pqty{\Momx{\gamma}+\Delta\Momx{\gamma}_{c,j}} - f(\Momx{\gamma})} \d R_{c,j} .
    \label{eq:SDE_Ito}
\end{equation}{}
For instance, eq.~\ref{eq:SDE_Ito} can be used to evaluate the stochastic dynamics of the square $(\Momx{\gamma})^2$ of a population moment.
An SDE for functions of more than one population moments, such as a cross-product  $f(\Momx{\gamma'},\Momx{\gamma''})=\Momx{\gamma'}\Momx{\gamma''}$,
can be obtained similarly, by evaluating the overall change before and after the transition due to the simultaneous update of both moments.

\section{Multivariate Gamma moment closure for compartment-population models}
Similarly to all moment-based approaches, also our moment equation method for compartment-population dynamics usually requires the use of some moment closure approximation in order to obtain a finite set of ODEs. 
A moment closure is a technique to approximate all occurrences of moments higher then a certain order with functions of lower order moments within a set of moment equations.
The accuracy of moment-closure depends on the particular model under consideration as well as its parameters and initial conditions~\cite{Schnoerr2014}.

For the purposes of our work, we require a multivariate closure scheme.
We compared three established multivariate closures across our case studies, i.e., the normal, lognormal and Gamma closure.
We found the Gamma closure proposed by~\cite{Lakatos2015_MultivariateClosure} to consistently outperform the other two closures, which is why we adopted this choice for all our case studies. 
We generally consider second-order closures (i.e., replacing moments of order higher than two), unless stated differently. In particular, we applied the following Gamma closure schemes:

\begin{itemize}
\item Closing a third-order cross-product between a moment $\Momx{\gamma}$ and a moment $\Momx{\xi}$, with $\Momx{\gamma}$ appearing to squared power
\begin{equation}
\Et{(\Momx{\gamma})^2\Momx{\xi}} = 2\frac{\Et{(\Momx{\gamma})^2}\Et{\Momx{\gamma}\Momx{\xi}}}{\Et{\Momx{\gamma}}} - \Et{(\Momx{\gamma})^2}\Et{\Momx{\xi}} .
\end{equation}

\item Closing a single moment appearing at order $3$, e.g. $\Et{\Momx{3}}=\sum_{x=0}^\infty x^3 \Et{\nx{x}}$
\begin{equation}
\Et{\Momx{3}} = 2\frac{\Et{\Momx{2}}^2}{\Et{\Momx{1}}} - \frac{\Et{\Momx{1}}\Et{\Momx{2}}}{\Et{N}} .
\end{equation}
which is also a Gamma closure, but corrected with the fact that the expected content distribution $\Et{\nx{x}}$ has normalization equal to $\Et{N}$.

\end{itemize}

\section{CASE STUDY: Nested birth-death process}
The transition classes that define the first case study are given by
\begin{align}
    \emptyset &\xrightharpoonup{h_I(\vn;y)} [y] && h_I(\vn;y)=k_I\pi_{Poiss}(y;\lambda) \notag \\
    [x]   &\xrightharpoonup{h_E(\vn;x)} \emptyset && h_E(\vn;x)=k_E\nx{x} \notag \\
    [x] &\xrightharpoonup{h_b(\vn;x)} [x+1] && h_b(\vn;x)=k_b \nx{x}  \notag \\
    [x] &\xrightharpoonup{h_d(\vn;x)} [x-1] && h_d(\vn;x)=k_d x \nx{x},
    \label{SCN:BirthDeath}
\end{align}
We proceed below with a step-by-step derivation of both SDE and moment equations for this model.

\subsection{Moment equations}
The expected trajectory for $\Et{N}$ is already reported in the main text. To characterize the dynamics of $N^2$, we consider again the SDE for $N$, which reads
\begin{equation}
	\d N = \d R_I - \d R_E
	\label{eq:SDE_N_BirthDeath}
\end{equation}
and we calculate the SDEs for $N^2$ using Ito's rule for counting processes
\begin{align}
\d N^2 &= \bqty{(N+1)^2-N^2}\d R_I + \bqty{(N-1)^2-N^2}\d R_E \notag \\
&= \pqty{1+2N}\d R_I + \pqty{1-2N}\d R_E .
\label{eq:SDE_N2_BirthDeath}
\end{align}{}
The expectation of~\ref{eq:SDE_N2_BirthDeath} is
\begin{equation}
\frac{\d \Et{N^2}}{\d t} = k_I\pqty{1+2\Et{N}} + k_E\pqty{\Et{N}-2\Et{N^2}}  .
\label{eq:ODE_N2_BirthDeath}
\end{equation}
We note that the evolution of $\Et{N^2}$ is already in closed form. 
The SDE for the total mass $\Momx{1}$ includes also the contribution of the birth-death reactions taking place in each compartment, thus
\begin{equation}
\d\Momx{1} = \sum_{y=0}^\infty (+y)\d R_{I,y} + \sum_{x=0}^\infty (-x) \d R_{E,x} + \sum_{x=0}^\infty (+1) \d R_{b,x} +\sum_{x=0}^\infty (-1) \d R_{d,x} .
\end{equation}
Recalling that the content of newly created compartments is chosen to be Poisson distributed with parameter $\lambda$, the expected trajectory for the total mass is
\begin{align}
\frac{\d\Et{\Momx{1}}}{\d t} &= k_I \Etb{\sum_{y=0}^\infty y \pi_I(y;\lambda)} - k_E \Etb{\sum_{x=0}^\infty x \nx{x}} 
+ k_b \Etb{\sum_{x=0}^\infty \nx{x}} - k_d \Etb{\sum_{x=0}^\infty x \nx{x}} \notag \\
&= k_I\lambda - k_E\Et{\Momx{1}} + k_b\Et{N} - k_d\Et{\Momx{1}} ,
\end{align}
as shown in the main paper. Similarly to $N^2$, we derive now the SDE for $(\Momx{1})^2$ by Ito's rule
\begin{align}
\d(\Momx{1})^2 =& \sum_{y=0}^\infty \bqty{(\Momx{1}+y)^2-(\Momx{1})^2} \d R_{I,y} + \sum_{x=0}^\infty \bqty{(\Momx{1}-x)^2-(\Momx{1})^2} \d R_{E,x} \notag \\
&+ \sum_{x=0}^\infty \bqty{(\Momx{1}+1)^2-(\Momx{1})^2} \d R_{b,x} + \sum_{x=0}^\infty \bqty{(\Momx{1}-1)^2-(\Momx{1})^2} \d R_{d,x} \notag \\
=& \sum_{y=0}^\infty \bqty{y^2+2y\Momx{1}} \d R_{I,y} + \sum_{x=0}^\infty \bqty{x^2-2x\Momx{1}} \d R_{E,x} \notag \\
&+ \sum_{x=0}^\infty \bqty{1+2\Momx{1}} \d R_{b,x} + \sum_{x=0}^\infty \bqty{1-2\Momx{1}} \d R_{d,x} .
\label{eq:SDE_M2_BirthDeath}
\end{align}
The expected trajectory for $\Et{(\Momx{1})^2}$ equals
\begin{align}
\frac{\d\Et{(\Momx{1})^2}}{\d t} =& \, k_I \Etb{\sum_{y=0}^\infty (y^2+2y\Momx{1}) \pi_I(y;\lambda)} + k_E \Etb{\sum_{x=0}^\infty (x^2 -2x\Momx{1})\nx{x}}  \notag \\
&+ k_b \Etb{\sum_{x=0}^\infty (1+2\Momx{1}) \nx{x}} + k_d \Etb{\sum_{x=0}^\infty (1-2\Momx{1}) x\nx{x}} \notag \\
=& \, k_I\lambda(1+\lambda+2\Et{\Momx{1}}) + k_E(\Et{\Momx{2}}-2\Et{(\Momx{1})^2})  \notag \\
&+ k_b(\Et{N}+2\Et{N\Momx{1}}) + k_d(\Et{\Momx{1}}-2\Et{(\Momx{1})^2}) .
\end{align}
This result introduces a dependency on two higher order moments, $\Et{\Momx{2}}$ and $\Et{N\Momx{1}}$. By observing that expectation of eq.~\ref{eq:SDE_Ito} can be written as
\begin{equation}
    \frac{\d \Et{f(\Momx{\gamma}})}{\d t} = \sum_{c\in\Cset} \Et{ \sum_{j\in\Jset_c} \bqty{f\pqty{\Momx{\gamma}+\Delta\Momx{\gamma}_{c,j}} - f(\Momx{\gamma})} \prop_{c,j}(\vn)} ,
    \label{eq:Expectationf}
\end{equation}
we find respectively
\begin{align}
\frac{\d\Et{\Momx{2}}}{\d t} =& \, k_I \Etb{\sum_{y=0}^\infty (y^2) \pi_I(y;\lambda)} + k_E \Etb{\sum_{x=0}^\infty (-x^2)\nx{x}}  \notag \\
&+ k_b \Etb{\sum_{x=0}^\infty \bqty{(x+1)^2-x^2} \nx{x}} + k_d \Etb{\sum_{x=0}^\infty  \bqty{(x-1)^2-x^2} x\nx{x}} \notag \\
=& \, k_I\lambda(1+\lambda) - k_E\Et{\Momx{2}} + k_b(\Et{N}+2\Et{\Momx{1}}) + k_d(\Et{\Momx{1}}-2\Et{\Momx{2}}) 
\label{eq:M2_BD}
\end{align}
and
\begin{align}
\frac{\d\Et{N\Momx{1}}}{\d t} =& \, k_I \Etb{\sum_{y=0}^\infty \bqty{(N+1)(\Momx{1}+y)-N\Momx{1}} \pi_I(y;\lambda)}  \notag \\ 
&+ k_E \Etb{\sum_{x=0}^\infty \bqty{(N-1)(\Momx{1}-x)-N\Momx{1}}\nx{x}}  \notag \\
&+ k_b \Etb{\sum_{x=0}^\infty \bqty{(N(\Momx{1}+1)-N\Momx{1}} \nx{x}} \notag \\
&+ k_d \Etb{\sum_{x=0}^\infty  \bqty{(N(\Momx{1}-1)-N\Momx{1}} x\nx{x}} \notag \\
=& \, k_I\pqty{\lambda(1+\Et{N})+\Et{\Momx{1}}} + k_E(\Et{\Momx{1}}-2\Et{N\Momx{1}}) + k_b\Et{N^2} - k_d\Et{N\Momx{1}}.
\label{eq:NM_BD}
\end{align}
Both eqs.~\ref{eq:M2_BD} and~\ref{eq:NM_BD} do not involve any higher-order moments.
In summary, we obtained a closed system of $6$ differential equations
\begin{align}
\frac{\d \Et{N}}{\d t} =&\, k_I - k_E\Et{N} \notag \\
\frac{\d \Et{N^2}}{\d t} =&\, k_I\pqty{1+2\Et{N}} + k_E\pqty{\Et{N}-2\Et{N^2}} \notag \\
\frac{\d\Et{\Momx{1}}}{\d t} =&\, k_I\lambda - k_E\Et{\Momx{1}} + k_b\Et{N} - k_d\Et{\Momx{1}} \notag \\
\frac{\d\Et{(\Momx{1})^2}}{\d t} =&\, k_I\lambda(1+\lambda+2\Et{\Momx{1}}) + k_E(\Et{\Momx{2}}-2\Et{(\Momx{1})^2}) \notag \\
&+ k_b(\Et{N}+2\Et{N\Momx{1}}) + k_d(\Et{\Momx{1}}-2\Et{(\Momx{1})^2}) \notag \\
\frac{\d\Et{\Momx{2}}}{\d t} =&\, k_I\lambda(1+\lambda) - k_E\Et{\Momx{2}} + k_b(\Et{N}+2\Et{\Momx{1}}) + k_d(\Et{\Momx{1}}-2\Et{\Momx{2}}) \notag \\
\frac{\d\Et{N\Momx{1}}}{\d t} =& \, k_I\pqty{\lambda(1+\Et{N})+\Et{\Momx{1}}} + k_E(\Et{\Momx{1}}-2\Et{N\Momx{1}}) + k_b\Et{N^2} - k_d\Et{N\Momx{1}}.
\label{eq:BD_ODEs}
\end{align}
After numerical integration, the time evolution of the standard deviations for $N$ and $\Momx{1}$ can be calculated by $\sqrt{\Et{N^2}-\Et{N}^2}$ and $\sqrt{\Et{(\Momx{1})^2}-\Et{\Momx{1}}^2}$, respectively.

\subsection{Steady-state properties}
As already explained in the main paper, in this model the dynamics of the total number of compartments $N$ is independent of other moments of the system. This can be observed from the SDE~\ref{eq:SDE_N_BirthDeath}, which corresponds to the dynamics of a birth-death process with propensities $\Prop_I(\vn)=k_I$ and $\Prop_E(\vn)=k_E N$, respectively. This shall not be confused with the chemical birth-death process of rates $k_b$ and $k_d$, affecting the content $x$ inside each compartment. Eq.~\ref{eq:SDE_N_BirthDeath} implies that the steady state number of compartments $N_\infty=\lim_{t\rightarrow\infty}N$ is Poisson distributed with mean $k_I/k_E$. This is resembled also by the steady-state solution of the moment equations, for which we find that $\Et{N_\infty} = k_I/k_E$ and $\Et{N^2_\infty}=k_I/k_E(1+k_I/k_E)$, so that 
\begin{equation}
\mathrm{Var}(N_\infty)=\Et{N^2_\infty}-\Et{N_\infty}^2 = k_I/k_E = \Et{N_\infty} ,
\label{eq:N_Poisson}
\end{equation}
consistent with Poisson noise, i.e. $N_\infty \sim \mathrm{Poisson}(k_I/k_E)$. 
From the steady-state solution of the moment equations we can further derive the steady-state total mass 
\begin{align}
    \Et{\Momx{1}_\infty} =  \frac{k_bk_I/k_E+k_I\lambda}{k_d+k_E} &= \frac{k_I}{k_E} \bqty{\frac{k_b}{k_d}\frac{1+\alpha\beta}{1+\alpha}} \notag \\ 
    &=\Et{N_\infty}\bqty{\frac{k_b}{k_d}\frac{1+\alpha\beta}{1+\alpha}},
    \label{eq:M_infty}
\end{align}
where we introduced the dimensionless parameters $\alpha=k_E/k_d$ and $\beta=\lambda/(k_b/k_d)$ .
We notice that the term in square brackets in eq.~\ref{eq:M_infty} is an estimate of the steady-state content per compartment, which we denote by $\Et{X_\infty} = \Et{\Momx{1}_\infty}/\Et{N_\infty}$.
Similarly, it is possible to calculate $\Et{\Momx{2}_\infty}$ and consequently identify $\Et{X^2_\infty}$. 
After some algebraic manipulations, we can obtain the variance-to-mean ratio of the steady-state content per compartment as follows
\begin{align}
	\frac{\mathrm{Var}(X_\infty)}{\Et{X_\infty}}&= \frac{\Et{X^2_\infty}-\Et{X_\infty}^2}{\Et{X_\infty}} \notag \\
	&= 1 + \frac{k_b}{k_d}\frac{\alpha}{(1+\alpha)(2+\alpha)}\frac{(\beta-1)^2}{1+\alpha\beta}  \notag \\
	&= 1 + \Et{X_\infty}\frac{\alpha}{2+\alpha}\frac{(\beta-1)^2}{(1+\alpha\beta)^2}
\label{eq:VMR_BirthDeath}
\end{align}
An interesting insight from this result is that the variance-to-mean ratio achieves the minimum value $1$ for $\beta=1$. This indicates that the steady-state compartment content follows Poissonian statistics if $\lambda= k_b/k_d$, meaning that the content of compartments entering the system matches in distribution the chemical birth-death process occurring inside each compartment. Instead, whenever $\lambda \neq k_b/k_d$, the compartment content will have a limiting distribution with super-Poissonian noise.

\begin{figure}[h]
\centering
\includegraphics[width=.8\linewidth]{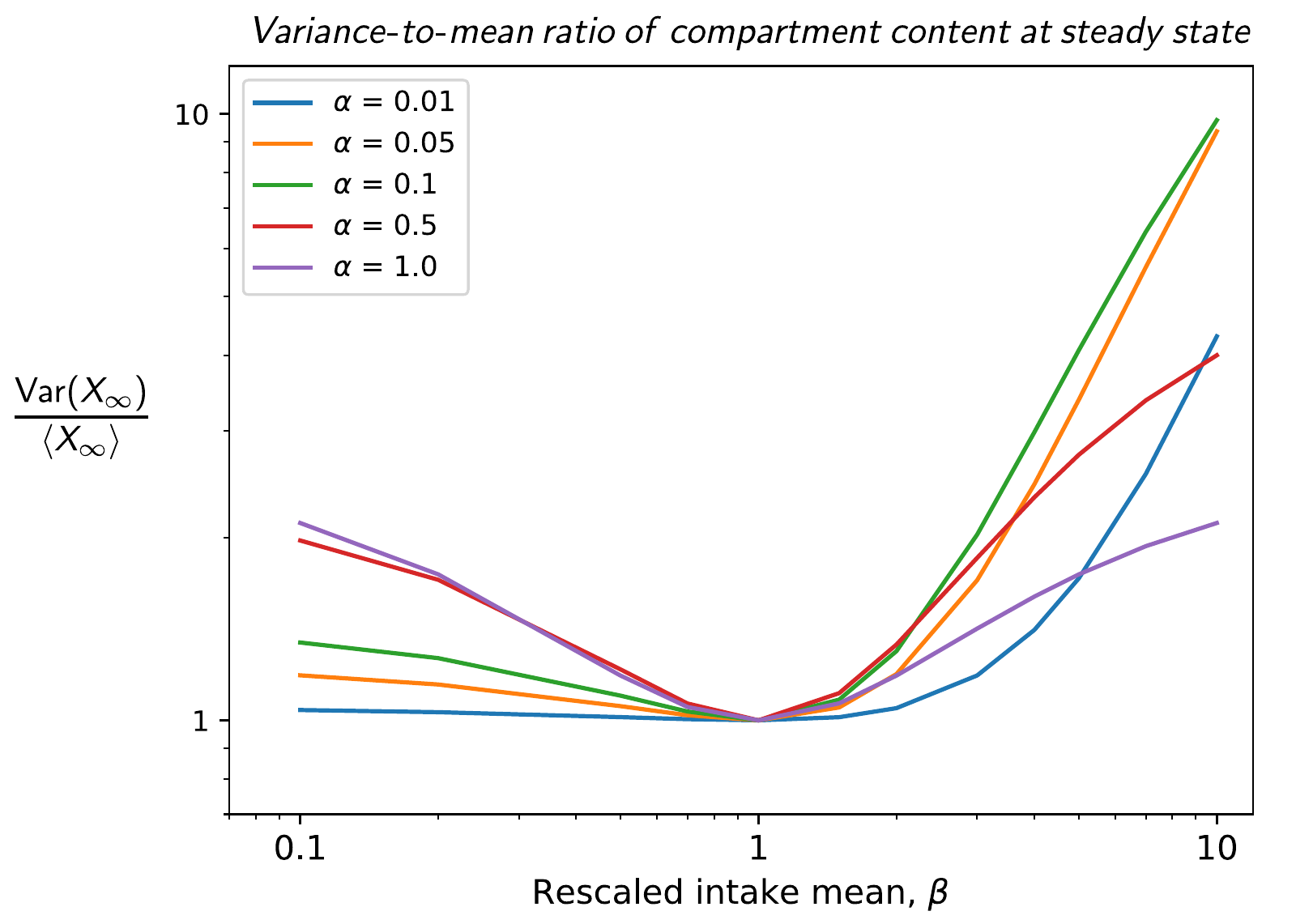}
\caption{The analytical result~\ref{eq:VMR_BirthDeath} is plotted as a function of $\beta=\lambda / (k_b/k_d)$, for different values of $\alpha=k_E/k_d$.}
\label{fig:VRM_Poisson}
\end{figure}

\subsection{Simulation parameters}
Figures 2B-C in the main paper have been generated using the following choice of parameters 
\begin{align}
k_I&=1 && \lambda=10 && k_E=0.01 && k_b=1 &&  k_d=0.1  \notag
\end{align}
and with initial condition 
\begin{equation*} 
\vn_0 : \n_0(1)=1, \,\,\,  n_0(x)=0 \,\,\,\forall x\neq 1
\end{equation*}
which represents a population comprising one compartment with content $x=1$.
Monte Carlo estimates have been obtained by averaging the outputs of $10^3$ stochastic simulations.

\section{CASE STUDY: Stochastic coagulation-fragmentation dynamics}
This case study is defined by the transition classes
\begin{align}
    \emptyset &\xrightharpoonup{h_I(\vn;y)} [y] && h_I(\vn;y)=k_I\pi_{Poiss}(y;\lambda) \notag \\
    [x]   &\xrightharpoonup{h_E(\vn;x)} \emptyset && h_E(\vn;x)=k_E\nx{x} \notag \\
    [x] + [x'] &\xrightharpoonup{h_C(\vn;x,x')} [x+x'] && h_C(\vn;x,x')=k_C \frac{n(x)(n(x')-\delta_{x,x'})}{1+\delta_{x,x'}} \notag \\
    [x] &\xrightharpoonup{h_F(\vn;x,y)} [y]+[x-y] && h_F(\vn;x,y)=k_F x \nx{x} \pi_F(y | x) .
    \label{SCN:CF}
\end{align}
Since the first two classes are equivalent to the previous case study, we will focus explicitly on the treatment of the coagulation and fragmentation transitions.

\subsection{Moment equations}

The total propensity of the random-coagulation class is found to be
\begin{equation}
    \Prop_C(\vn)=\sum_{x=0}^\infty \sum_{x'=0}^\infty k_C\frac{\nx{x}\pqty{\nx{x'}-\delta_{x,x'}}}{2}=k_C\frac{N(N-1)}{2} ,
    \label{eq:PropHc}
\end{equation}
where the factor $2$ accounts for double counting in the case $x\neq x'$ and arises from the binomial $\binom{\nx{x}}{2}$ for $x=x'$. We find that the total propensity is equal to the coagulation rate $k_C$ multiplied by the number of distinct compartment pairs $\binom{N}{2}$, which is consistent with a random coagulation mechanism.

As reported in Table 1 in the main text, the fragmentation transition class is in general defined through
\begin{equation}
[\vb{x}] \xrightharpoonup{\prop_F(\vn;\vb{x},\vb{y})} [\vb{y}] + [\vb{x} - \vb{y}] ,
\end{equation}
where we have already substituted the mass-conservation constraint in the right-hand side.
The corresponding propensity function is
\begin{equation}
\prop_F(\vn;\vb{x},\vb{y})=k_F g_F(\vb{x}) \nx{\vb{x}} \pi_F(\vb{y}|\vb{x}).
\end{equation}
Note that $ g_F(\vb{x})$  controls the probability of the "mother" compartment to undergo a fragmentation event. Next, the normalized probability distribution $\pi_F(\vb{y}|\vb{x})$  expresses the probability that one of the two fragments has a content equal to $\vb{y}$. 
$\pi_F(\vb{y}|\vb{x})$ must be symmetric with respect to $\vb{y}$ and $\vb{x}-\vb{y}$.
In this example, the compartment content is one-dimensional (i.e. $\vb{x} = x$) and we chose $g_F(x)=x$ and for $\pi_F(\vb{y}|\vb{x})$ a uniform fragment distribution
\begin{equation}
    \pi_F(y|x)=\frac{1}{x+1}   \,\,\,\,\, \text{with} \,\,\,\,\, 0 \leq y \leq x ,
    \label{eq:UnifFrag}
\end{equation}
which permits the creation of empty daughter compartments too.
The corresponding total fragmentation propensity is equal to
\begin{align}
    \Prop_F(\vn)&= \sum_{x=0}^\infty \sum_{y=0}^x k_Fg_F(x)\nx{x} \pi_F(y|x)\notag \\
    &= k_F  \sum_{x=0}^\infty  x \nx{x} \sum_{y=0}^x\pi_F(y|x) =k_F\Momx{1} .
\end{align}
The moment equations for this model read
\begin{align}
\frac{\d \Et{N}}{\d t} =&\, k_I - k_E\Et{N} - \frac{k_C}{2}\pqty{\Et{N^2}-\Et{N}} +k_F\Et{\Momx{1}} \notag \\
\frac{\d \Et{N^2}}{\d t} =&\, k_I\pqty{1+2\Et{N}} + k_E\pqty{\Et{N}-2\Et{N^2}} + \frac{k_C}{2}\pqty{\Et{N^2}-\Et{N}} -k_C\pqty{\Et{N^3}-\Et{N^2}} \notag \\
&+ k_F\pqty{\Et{\Momx{1}}+2\Et{N\Momx{1}}} \notag \\
\frac{\d\Et{\Momx{1}}}{\d t} =&\, k_I\lambda - k_E\Et{\Momx{1}} \notag \\
\frac{\d\Et{(\Momx{1})^2}}{\d t} =&\, k_I\lambda(1+\lambda+2\Et{\Momx{1}}) + k_E(\Et{\Momx{2}}-2\Et{(\Momx{1})^2}) \notag \\
\frac{\d\Et{\Momx{2}}}{\d t} =&\, k_I\lambda(1+\lambda) - k_E\Et{\Momx{2}} +k_C\pqty{\Et{(\Momx{1})^2}-\Et{\Momx{2}}} +\frac{k_F}{3}\pqty{\Et{\Momx{2}}-\Et{\Momx{3}}} \notag \\
\frac{\d\Et{N\Momx{1}}}{\d t} =& \, k_I\pqty{\lambda(1+\Et{N})+\Et{\Momx{1}}} + k_E(\Et{\Momx{1}}-2\Et{N\Momx{1}}) \notag \\
 &+ \frac{k_C}{2}\pqty{\Et{N\Momx{1}}-\Et{N^2\Momx{1}}} +k_F\Et{(\Momx{1})^2}
 \label{eq:MomEq_CF}
\end{align}
and we employ the Gamma closure scheme on the following third-order moments
\begin{align}
&\Et{N^3} =  2\frac{\Et{N^2}^2}{\Et{N}} - \Et{N^2}\Et{N} \notag \\
&\Et{N^2\Momx{1}} = 2\frac{\Et{N^2}\Et{N\Momx{1}}}{\Et{N}} - \Et{N^2}\Et{\Momx{1}} \notag \\
&\Et{\Momx{3}} =  2\frac{\Et{(\Momx{2}})^2}{\Et{\Momx{1}}} - \frac{\Et{\Momx{2}}\Et{\Momx{1}}}{\Et{N}} .
\label{eq:Gamma_CF}
\end{align}
Since the total mass $\Momx{1}$ is not affected by coagulation or fragmentation events, we find the equations for $\Et{\Momx{1}}$ and $\Et{(\Momx{1})^2}$ to be identical to the previous case study in terms of intake and exit contributions. 
The derivation of the equations for $\Et{N}$ and $\Et{N^2}$ is also straightforward and resembles closely the treatment of eqs.~\ref{eq:SDE_N2_BirthDeath} and~\ref{eq:ODE_N2_BirthDeath}. Here we explicitly show only the derivation of the coagulation-fragmentation contributions for $\frac{\d\Et{\Momx{2}}}{\d t}$ 
\begin{align}
\frac{\d\Et{\Momx{2}}}{\d t}\biggr\rvert_{C,F}  =& \, \Etb{ \sum_{x=0}^\infty \sum_{x'=0}^\infty \bqty{(x+x')^2-(x)^2-(x')^2} k_C\frac{\nx{x}\pqty{\nx{x'}-\delta_{x,x'}}}{2} } \notag \\
&+ \Etb{\sum_{x=0}^\infty \sum_{y=0}^x \bqty{y^2+(x-y)^2-x^2} k_F x \nx{x} \pi_F(y | x) } \notag \\
 =& \, k_C \Etb{ \sum_{x=0}^\infty \sum_{x'=0}^\infty \bqty{2xx'} \frac{\nx{x}\pqty{\nx{x'}-\delta_{x,x'}}}{2} } \notag \\
&+ k_F \Etb{\sum_{x=0}^\infty x \nx{x} \sum_{y=0}^x 2\bqty{y^2-xy} \pi_F(y | x) } \notag \\
=& \, k_C \Etb{ \sum_{x=0}^\infty x \nx{x} \sum_{x'=0}^\infty x' \nx{x'} } -k_C\Etb{ \sum_{x=0}^\infty x^2 \nx{x}}  \notag \\
&+ 2 k_F \Etb{\sum_{x=0}^\infty x\nx{x} \bqty{\frac{2x^2+x}{6} -x\frac{x}{2}}} \notag \\
=& \, k_C\pqty{\Et{(\Momx{1})^2}-\Et{\Momx{2}}} +\frac{k_F}{3}\pqty{\Et{\Momx{2}}-\Et{\Momx{3}}} .
\end{align}

\begin{figure}[h!]
\centering
\includegraphics[width=0.7\linewidth]{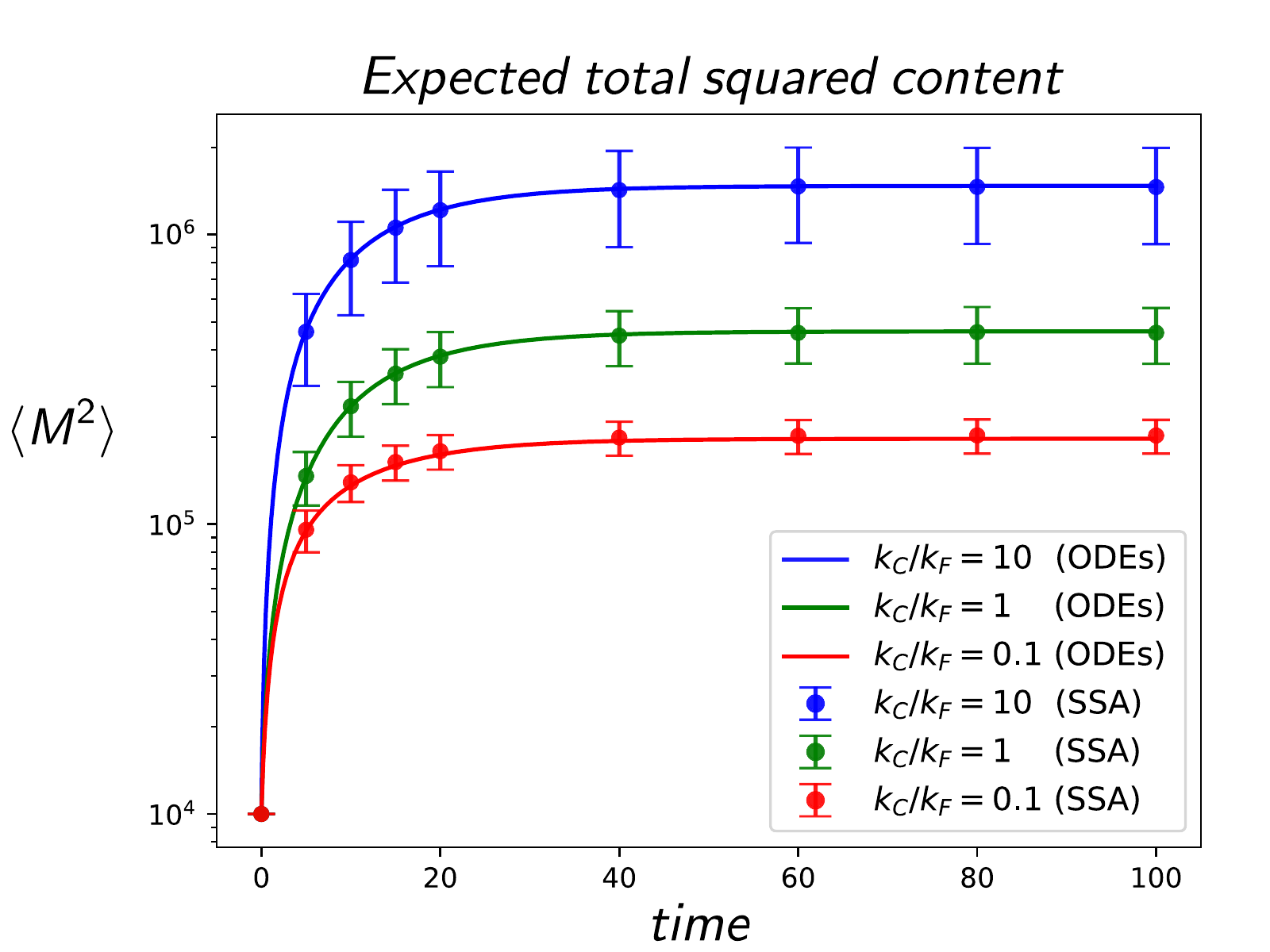}
\caption{Agreement of the expected dynamics of the second order moment $\Momx{2}=\sum_{x=0}^\infty x^2\nx{x}$ between the Gamma-closed moment equations (ODEs) and the estimate obtained from stochastic simulations (SSA). Dots and error bars represent the mean value and one standard deviation above and below the mean. The variability of $\Et{\Momx{2}}$ cannot be reported from the solution of the moment equations because the equation for $\Et{(\Momx{2})^2}$ was not included.}
\label{fig:M2_CF}
\end{figure}

\subsection{Simulation parameters}
In Figures 2E-F in the main paper we used the following choice of parameters
\begin{align}
k_I&=10 && \lambda=50 && k_E=0.1  &&  k_F=5\cdot 10^{-3} \notag
\end{align}
and $k_C$ takes the values $[0.1k_F, k_F, 10k_F]$ in the three cases. 
The initial condition was set to
\begin{equation*} 
\vn_0 : \n_0(10)=100, \,\,\,  n_0(x)=0 \,\,\,\forall x\neq 1
\end{equation*}
which represents a population comprising $100$ compartment with content $x=10$.
Monte Carlo estimates have been obtained by averaging the outputs of $10^4$ stochastic simulations, for each parameter combination.
In Fig.~\ref{fig:M2_CF} we show the accuracy of the solution for $\Et{\Momx{2}}$ obtained in these settings, next to the estimate from exact stochastic simulations.

\section{CASE STUDY: Transcription dynamics in a cell community}
The state space of the compartment content $(x_G,x_S)$ is $\SpaceX = [0,1] \times \Nat$, where $x_G=1$ represent the binary gene variable in active state. 
The transcription network in each cell is modelled as 
\begin{align}
   & [x_G,x_S] \xrightharpoonup{h_b^G(\vn;\vb{x})} [1,x_S] && h_b^G(\vn;\vb{x})=k_b^G (1-x_G) \nx{\vb{x}} \notag \\
   & [x_G,x_S] \xrightharpoonup{h_d^G(\vn;\vb{x})} [0,x_S] && h_d^G(\vn;\vb{x})=k_d^G x_G \nx{\vb{x}} \notag \\
   & [x_G,x_S] \xrightharpoonup{h_S(\vn;\vb{x})} [x_G,x_S+1] && h_S(\vn;\vb{x})=k_S x_G \nx{\vb{x}}  \notag \\
   & [x_G,x_S] \xrightharpoonup{h_b^S(\vn;\vb{x})} [x_G,x_S+1] && h_b^S(\vn;\vb{x})=k_b^S \nx{\vb{x}} \notag \\
   & [x_G,x_S] \xrightharpoonup{h_d^S(\vn;\vb{x})} [x_G,x_S-1] && h_d^S(\vn;\vb{x})=k_d^S x_S\nx{\vb{x}} \notag \\ 
    \label{SCN:ReactionsComm}
\end{align}
The communication class for which active cells can induce activation in inactive cells is defined by
\begin{equation}
	[\vb{x}] + [\vb{x'}] \xrightharpoonup{\prop_{com}(\vn;\vb{x},\vb{x'})} [1,x_S] + [1,x_S']
	\label{eq:CommEvent}
\end{equation}
where
\begin{align}
	&\prop_{com}(\vn;\vb{x},\vb{x'}) = k_{com} g_{com}(\vb{x},\vb{x'})\frac{\nx{\vb{x}}(\nx{\vb{x}'}-\delta_{\vb{x},\vb{x}'})}{1+\delta_{\vb{x},\vb{x}'}} \notag \\
	& \text{with   } g_{com}(\vb{x},\vb{x'})=x_G(1-x_G')+x_G'(1-x_G) .
	\label{eq:CommProp}
\end{align}
The content-dependent function $g_{com}$ ensures the transition to happen only between an active and inactive cell. 
This expression for $g_{com}$ also implies that the cell-cell activation happens with equal probability among any active or inactive cell, irrespectively of their protein content, since there is no explicit dependency on $x_S$ and $x_S'$.

\subsection{Moment equations}
The total propensity of the cell communication class is found to be
\begin{align}
	\Prop_{com}(\vn) =&\, \sum_{\vb{x}} \sum_{\vb{x}'} k_{com} \pqty{x_G(1-x_G')+x_G'(1-x_G)} \frac{\nx{\vb{x}}\pqty{\nx{\vb{x}'}-\delta_{\vb{x},\vb{x}'}}}{2} \notag \\
	=&\,  \frac{k_{com}}{2} \sum_{\vb{x}} \sum_{\vb{x}'} \pqty{x_G+x_G'-2x_Gx_G')} \nx{\vb{x}}\nx{\vb{x'}}
	   \,- \frac{k_{com}}{2} \sum_{\vb{x}}  \pqty{2x_G-2x_G^2} \nx{\vb{x}} \notag \\
	=&\,  \frac{k_{com}}{2} \pqty{\sum_{\vb{x}} x_G\nx{\vb{x}}\sum_{\vb{x}'}\nx{\vb{x}'} + \sum_{\vb{x}} \nx{\vb{x}}\sum_{\vb{x}'}x_G'\nx{\vb{x}'}}
	\,\notag \\ &-k_{com}\sum_{\vb{x}} x_G\nx{\vb{x}}\sum_{\vb{x}'}x_G'\nx{\vb{x}'}  \notag \\
	=&\, k_{com} \pqty{N\Momx{1,0}-(\Momx{1,0})^2} \notag \\
	=&\, k_{com} \Momx{1,0}\pqty{N_0-\Momx{1,0}},
\end{align}
where from the second to third line we have made use of the fact that $x_G=x_G^2$, since $x_G \in [0,1]$. 
The factor $1/2$ under the propensity function originates in the same way as in the computation~\ref{eq:PropHc}.
As expected, the total communication propensity equals the rate $k_{com}$ times the product of the number of active and inactive cells.

The system of moment equations describing the evolution of this model is
\begin{align}
\frac{\d\Et{\Momx{1,0}}}{\d t} =&\, k_{com}\pqty{N_0\Et{\Momx{1,0}} - \Et{(\Momx{1,0})^2}} +k_b^G\pqty{N_0-\Et{\Momx{1,0}}} -k_d^G \Et{\Momx{1,0}} \notag \\
\frac{\d\Et{(\Momx{1,0})^2}}{\d t} =&\, k_{com}\pqty{N_0\Et{\Momx{1,0}} - \Et{(\Momx{1,0})^2}} +2k_{com}\pqty{N_0\Et{(\Momx{1,0})^2} - \Et{(\Momx{1,0})^3}} \notag \\
&+   k_b^G\bqty{N_0-\Et{\Momx{1,0}}+2\pqty{N_0\Et{\Momx{1,0}}-\Et{(\Momx{1,0})^2}}} \notag \\  &+k_d^G \pqty{\Et{\Momx{1,0}} -2\Et{(\Momx{1,0})^2} } \notag \\ 
\frac{\d\Et{\Momx{0,1}}}{\d t} =&\, k_S\Et{\Momx{1,0}} + k_b^SN_0 -k_d^S \Et{\Momx{0,1}} \notag \\
\frac{\d\Et{(\Momx{0,1})^2}}{\d t} =&\, k_S\pqty{\Et{\Momx{1,0}}+2\Et{\Momx{1,0}\Momx{0,1}}} + k_b^SN_0\pqty{1+2\Et{\Momx{0,1}}} \notag \\ 
&+k_d^S \pqty{\Et{\Momx{0,1}} -2\Et{(\Momx{0,1})^2}} \notag \\
\frac{\d\Et{\Momx{1,0}\Momx{0,1}}}{\d t} =& \, k_{com}\pqty{N_0\Et{\Momx{1,0}\Momx{0,1}}-\Et{(\Momx{1,0})^2\Momx{0,1}}} + k_S\Et{(\Momx{1,0})^2} +k_b^S N_0\Et{\Momx{1,0}} \notag \\
 &+ k_b^G\pqty{N_0\Et{\Momx{0,1}}-\Et{\Momx{1,0}\Momx{0,1}}} -\pqty{k_d^G+k_d^S}\Et{\Momx{1,0}\Momx{0,1}}
\end{align}
and we employ the Gamma closure scheme on the following third-order moments:
\begin{align}
&\Et{(\Momx{1,0})^3} =  2\frac{\Et{(\Momx{1,0})^2}^2}{\Et{\Momx{1,0}}} - \Et{(\Momx{1,0})^2}\Et{\Momx{1,0}} \notag \\
&\Et{(\Momx{1,0})^2\Momx{0,1}} = 2\frac{\Et{(\Momx{1,0})^2}\Et{\Momx{1,0}\Momx{0,1}}}{\Et{\Momx{1,0}}} - \Et{(\Momx{1,0})^2}\Et{\Momx{0,1}}  .
\end{align}

\begin{figure}[h!]
\centering
\includegraphics[width=0.7\linewidth]{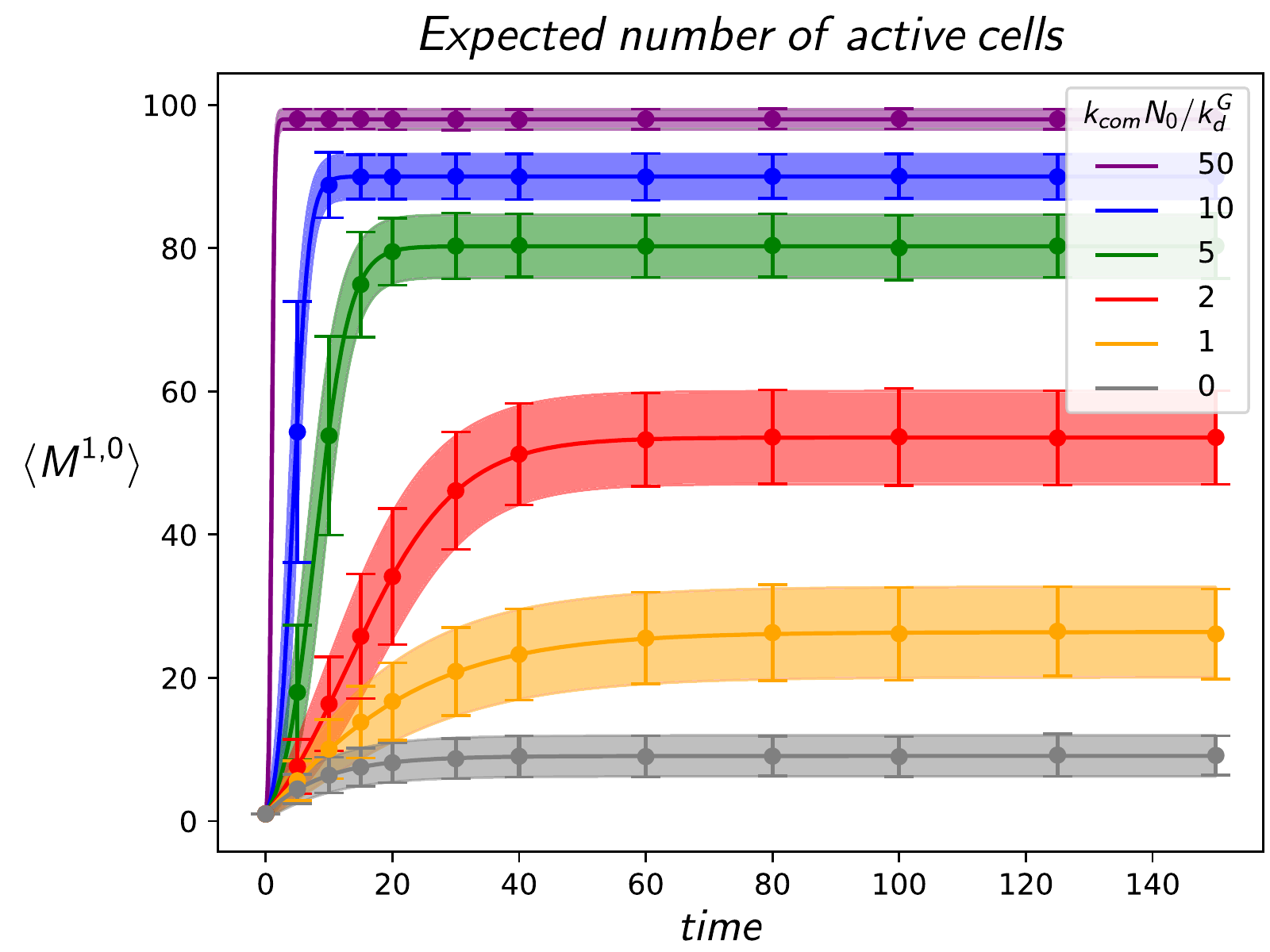}
\caption{Expected total number of active cells $\Momx{1,0}$ and its variability for different values of the communication rate. Results from moment equations are shown in full lines and shaded areas, and estimates from stochastic simulation in dots and error bars. The mean value is surrounded above and below by one standard deviation.}
\label{fig:Comm_Protein}
\end{figure}

\subsection{Simulation parameters}
In Figures 3B-C in the main paper we used for the following choice of parameters 
\begin{align}
k_b^G&=0.01 && k_d^G=0.1 && k_S=1  &&  k_b^S=0.1 && k_d^S=0.05 \notag
\end{align}
and $k_{com}$ takes the values $[0, 10^{-3}, 2\cdot 10^{-3}, 5\cdot 10^{-3} , 10^{-2}, 5\cdot 10^{-2}]$. 
The initial condition was set to
\begin{equation*} 
\vn_0 : n_0([1,1])=1, \,\,\, n_0(\mathbf{0})=99 , 
\end{equation*}
which represents a population comprising $100$ cells, one of which is in active state and contains one protein molecule and the remaining $99$ are in inactive state and with empty content.
Monte Carlo estimates have been obtained by averaging the outputs of $10^3$ stochastic simulations, for each parameter combination.
In Fig.~\ref{fig:Comm_Protein} we report the expected cell activation dynamics across the cell population, which complements Fig. 3B of the main paper.

\section{CASE STUDY: Stem cell population dynamics}
Also in this case study, the state space of the compartment content $\vb{x}=(x_G,x_S)$ is $\SpaceX = [0,1] \times \Nat$, where $x_G=1$ indicates a stem cell and $x_G=0$ a differentiated cell. 
The considered transition classes are
\begin{align}
   & [\vb{x}] \xrightharpoonup{h_F^+(\vn;\vb{x})} [x_G,0] + [x_G,0] && h_F^+(\vn;\vb{x})=k_F^+ x_G x_S \nx{\vb{x}} \notag \\
   & [\vb{x}] \xrightharpoonup{h_F^-(\vn;\vb{x})} [x_G,0] + [1-x_G,0] && h_F^-(\vn;\vb{x})=k_F^- x_G x_S \nx{\vb{x}} \notag \\
   & [\vb{x}] \xrightharpoonup{h_S(\vn;\vb{x})} [x_G,x_S+1] && h_S(\vn;\vb{x})=k_S x_G \nx{\vb{x}}  \notag \\
   & [\vb{x}]   \xrightharpoonup{h_E(\vn;\vb{x})} \emptyset && h_E(\vn;\vb{x})=k_E(1-x_G)\nx{\vb{x}} \notag \\
   & [\vb{x}] + [\vb{x}']  \xrightharpoonup{h_{nf}(\vn;\vb{x},\vb{x}',\vb{y},\vb{y'})} [\vb{y}] + [\vb{y'}] 
    \label{SCN:StemCell}
\end{align}
with
\begin{equation}
h_{nf}(\vn;\vb{x},\vb{x}',\vb{y},\vb{y'})=k_{nf} x_Gx_G' \frac{n(\vb{x})(n(\vb{x}')-\delta_{\vb{x},\vb{x}'})}{1+\delta_{\vb{x},\vb{x}'}}\pi_{nf}(\vb{y},\vb{y'}\mid\vb{x},\vb{x}')
\label{eq:FeedbackPropensity}
\end{equation}
and outcome distribution is given by
\begin{equation}
\pi_{nf}(\vb{y},\vb{y'}\mid\vb{x},\vb{x}') = \begin{cases}
\frac{1}{2} & \text{if \,} \vb{y}=\vb{x},\,y_G'=1-x_G',\,y_S'=x_S' \\
\frac{1}{2} & \text{if \,} \vb{y'}=\vb{x}',\,y_G=1-x_G,\,y_S=x_S .
\end{cases}
\end{equation}
The law implemented by the propensity of eq.~\ref{eq:FeedbackPropensity} can be understood as follows: first, two stem cells are randomly selected from the current state $\vn$, as enforced by the choice $g_{nf}(\vb{x},\vb{x}')=x_Gx_G'$. Then, one of the two stem cells is randomly differentiated by switching $x_G$ to zero, as encoded by $\pi_{nf}$. 
Note that, also for the remaining transition classes, the selectivity of the event on either stem cells or differentiated cells is achieved thanks to the $x_G$-dependence. For instance, the division events can occur only on stem cells, since their propensities $h_F^+$ and $h_F^-$ would be zero when the reactant cell has $x_G=0$, i.e. is a differentiated cell. The same is true for $h_S$. On the contrary, the death event applies only to differentiated cells, since $h_E(\vn;\vb{x})=0$ for stem cells.
The total class propensity for the feedback transition class is given by
\begin{align}
    \Prop_{nf}(\vn)&=\sum_{\vb{x},\vb{x}'} k_{nf} x_Gx_G'\frac{\nx{\vb{x}}\pqty{\nx{\vb{x}'}-\delta_{\vb{x},\vb{x}'}}}{2} \notag \\
    &= \frac{k_{nf}}{2}\pqty{\sum_{\vb{x}}x_G\nx{\vb{x}}\sum_{\vb{x}'}x_G'\nx{\vb{x}'}-\sum_{\vb{x}}x_G\nx{\vb{x}}} \notag \\
    &=k_{nf}\frac{\Momx{1,0}(\Momx{1,0}-1)}{2} = k_{nf} \binom{\Momx{1,0}}{2},
    \label{eq:PropStemCellFeedback}
\end{align}
where the binomial factor reflects that the negative feedback acts between randomly chosen pairs of stem cells.
In particular, the factor $1/2$ appearing in the first line of eq.~\ref{eq:PropStemCellFeedback} originates from $\pi_{nf}$ whenever $\vb{x} \neq \vb{x}'$ and from the binomial factor of eq.~\ref{eq:FeedbackPropensity} when $\vb{x}=\vb{x}'$, in which case $\pi_{nf}$ gives only one distinguishable outcome and thus contributes with value $1$.
The total propensities for the remaining transition classes of the model are readily computed:
\begin{align}
\Prop_F^+(\vn) &= k_F^+\Momx{1,1} \notag \\ 
\Prop_F^-(\vn) &= k_F^-\Momx{1,1} \notag \\ 
\Prop_S(\vn) &= k_S\Momx{1,0} \notag \\ 
\Prop_E(\vn) &= k_E\pqty{N-\Momx{1,0}}
\label{eq:PropStemCellOthers}
\end{align}

\subsection{Moment equations}
Due to the combined presence of the moment $\Momx{1,1}$ and the structure of the feedback class, the derivation of a closed set of moment equations is more challenging than for the previous models.
This is why it is important to focus on deriving equations for moments directly entering in the total class propensities, and at the same time to avoid further equations for less relevant population statistics which would introduce additional higher order dependencies.
From eq.~\ref{eq:PropStemCellFeedback} and~\ref{eq:PropStemCellOthers} we see that, together with $\Momx{1,1}$, the overall dynamics is governed by the total number of cells $N$ and the number of stem cells $\Momx{1,0}$. The SDEs for these moments are given by
\begin{align}
\d N &= \sum_{\vb{x}} \pqty{\d R_{F,\vb{x}}^+ + \d R_{F,\vb{x}}^- - \d R_{E,\vb{x}}} = \d R_{F}^+ + \d R_{F}^- - \d R_{E} \notag \\
\d \Momx{1,0} &= \sum_{\vb{x}} \d R_{F,\vb{x}}^+ - \sum_{\vb{x}}\sum_{\vb{x}'} \d R_{nf,\vb{x},\vb{x}'} = \d R_{F}^+ - \d R_{nf}\notag \\
\d \Momx{1,1} &= \sum_{\vb{x}} \d R_{S,\vb{x}} -\sum_{\vb{x}} x_S \pqty{\d R_{F,\vb{x}}^+ + \d R_{F,\vb{x}}^-} - \sum_{\vb{x}}\sum_{\vb{x}'} x_S\d R_{nf,\vb{x},\vb{x}'}  ,
\label{eq:SDEsStemCell}
\end{align}
where the enumeration of single transitions spans $\SpaceX$ for all transition classes with exception of the feedback, where all distinguishable transitions can be written as the iteration over $\vb{x},\vb{x'}\in\SpaceX^2$ and denoting with $\vb{x}$ the content of the cell which differentiates.
The first two equations have been rewritten also in terms of the global reaction counters, which better highlight their meaning.
We see how the total number of cells is increased by $1$ whenever a division event occurs and decreases by $1$ in occurrence of the death of any differentiated cell. On the other hand, the number of stem cells $\Momx{1,0}$ is increased by $1$ only when a symmetric division occurs, and is decreased by $1$ as a result of a feedback conversion. On the opposite, $\Momx{1,0}$ is left unaffected by asymmetric divisions.

After taking expectations on~\ref{eq:SDEsStemCell} and including further equations for some direct moment dependencies, we obtain the following set of moment equations
\begin{align}
\frac{\d \Et{N}}{\d t} =& \, \pqty{k_F^+ + k_F^-}\Et{\Momx{1,1}} -k_E\pqty{\Et{N}-\Et{\Momx{1,0}}} \notag \\
\frac{\d \Et{N^2}}{\d t} =& \, \pqty{k_F^+ + k_F^-}\pqty{\Et{\Momx{1,1}}+2\Et{N\Momx{1,1}}} +k_E\bqty{\Et{N}-\Et{\Momx{1,0}}-2\pqty{\Et{N^2}-\Et{N\Momx{1,0}}}} \notag \\
\frac{\d\Et{\Momx{1,0}}}{\d t} =&\, k_F^+ \Et{\Momx{1,1}} - k_{nf}\frac{\Et{(\Momx{1,0})^2}-\Et{\Momx{1,0}}}{2} \notag \\
\frac{\d\Et{(\Momx{1,0})^2}}{\d t} =&\, k_F^+ \pqty{\Et{\Momx{1,1}}+2\Et{\Momx{1,0}\Momx{1,1}}} \notag \\ &+ k_{nf}\frac{\Et{(\Momx{1,0})^2}-\Et{\Momx{1,0}}}{2} - k_{nf}\pqty{\Et{(\Momx{1,0})^3}-\Et{(\Momx{1,0})^2}} \notag \\
\frac{\d\Et{\Momx{1,1}}}{\d t} =&\, k_S\Et{\Momx{1,0}} -\pqty{k_F^+ + k_F^-}\Et{\Momx{1,2}} - k_{nf}\frac{\Et{\Momx{1,0}\Momx{1,1}}-\Et{\Momx{1,1}}}{2} \notag \\
\frac{\d\Et{\Momx{1,2}}}{\d t} =&\, k_S\pqty{\Et{\Momx{1,0}}+2\Et{\Momx{1,1}}} -\pqty{k_F^+ + k_F^-}\Et{\Momx{1,3}} - k_{nf}\frac{\Et{\Momx{1,0}\Momx{1,2}}-\Et{\Momx{1,2}}}{2}
\label{eq:MomEq_StemCells}
\end{align}
and we employ the Gamma closure scheme
\begin{align}
&\Et{(\Momx{1,0})^3} =  2\frac{\Et{(\Momx{1,0})^2}^2}{\Et{\Momx{1,0}}} - \Et{(\Momx{1,0})^2}\Et{\Momx{1,0}} \notag \\
&\Et{\Momx{1,3}} = 2\frac{\Et{(\Momx{1,2})^2}^2}{\Et{\Momx{1,1}}} - \frac{\Et{\Momx{1,2}}\Et{\Momx{1,1}}}{\Et{\Momx{1,0}}}  .
\end{align}
For the remaining cross-moments, we use a mean-field approximation. This is understood as the following substitution
\begin{equation}
\Et{\Momx{\gamma}\Momx{\xi}}=\Et{\Momx{\gamma}}\Et{\Momx{\xi}} ,
\end{equation}
which means to neglect the correlation between the two moments $\Momx{\gamma}$ and $\Momx{\xi}$.  
We applied the mean-field approximation on $\Et{N\Momx{1,0}}$, $\Et{N\Momx{1,1}}$, $\Et{\Momx{1,0}\Momx{1,1}}$ and $\Et{\Momx{1,0}\Momx{1,2}}$.
In this example, one of the effects of such simplification is noticeable from the expected fluctuations of $N$ and $\Momx{1,0}$ in Fig.~3F: since part of the noise contributing to the dynamics of $\Et{N^2}$ and $\Et{(\Momx{1,0})^2}$ is neglected, the variability predicted by moment equations is smaller than the error bars obtained by exact stochastic simulations.

\subsection{Simulation parameters}
\label{sec:reference_param_stem}
Figures 3E-F and the parameter reference for figures 3G-H in the main paper correspond to the following choice of parameters 
\begin{align}
k_F^+&=k_F^-=5\cdot 10^{-3} && k_S=10 &&   && k_E=0.05 && k_{nf}=0.01 \notag
\end{align}
The two initial conditions of Fig. 3F are
\begin{equation*} 
\vn_0 : n_0([1,1])=1, \,\,\,  n_0(\vb{x})=0 \,\,\,\forall \vb{x}\neq [1,1]
\end{equation*}
and
\begin{equation*} 
\vn_0 : n_0([1,1])=100, \,\,\,  n_0(\vb{x})=0 \,\,\,\forall \vb{x}\neq [1,1] ,
\end{equation*}
which represent a population made either of one or $100$ stem cells, each having $x_S=1$.
Monte Carlo estimates have been obtained by averaging the outputs of $10^3$ stochastic simulations.

\begin{figure}[h!]
\centering
\includegraphics[width=0.7\linewidth]{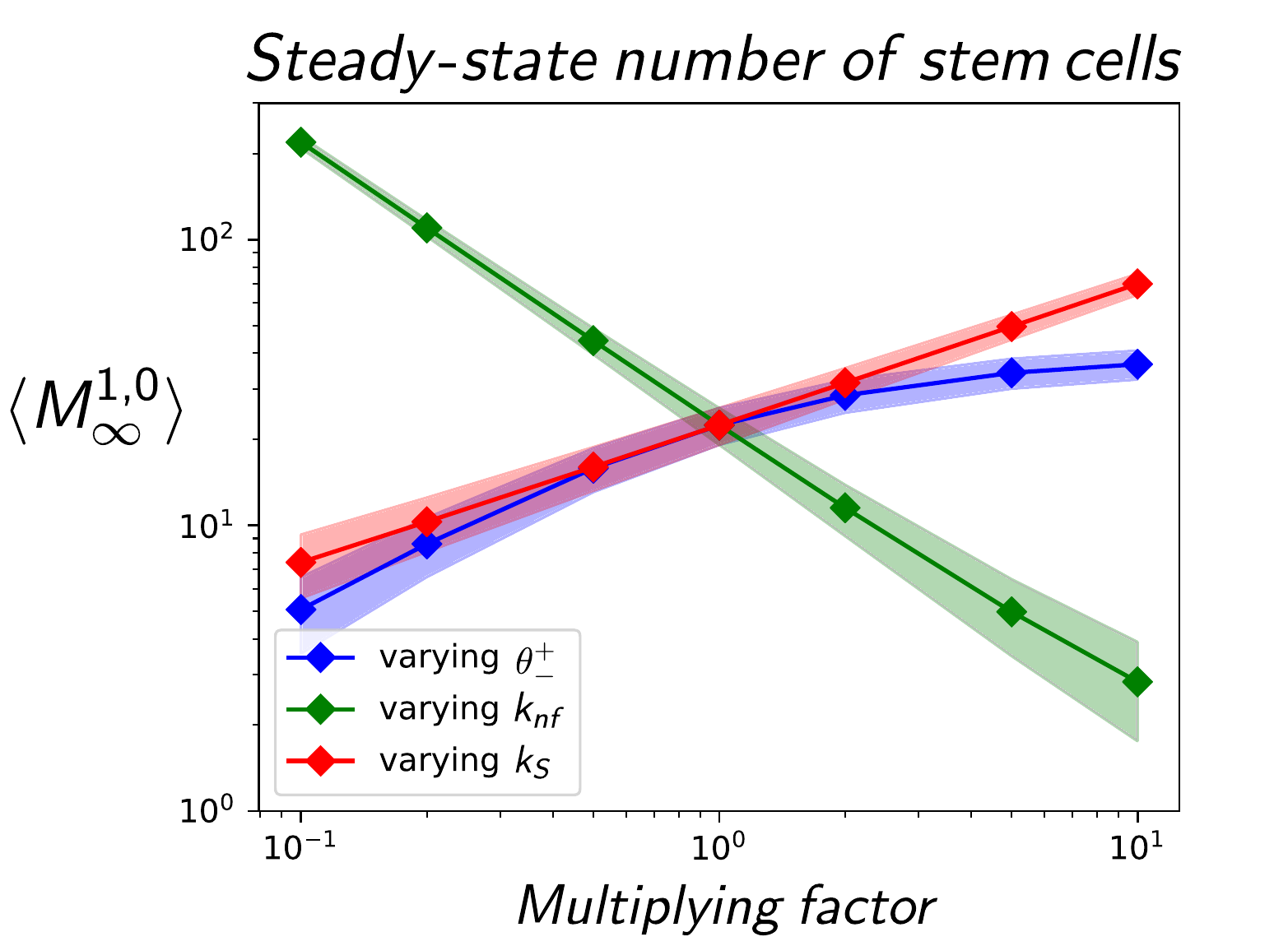}
\caption{Steady-state number of stem cells upon variations of different parameters, obtained from moment equations. The mean value is surrounded above and below by one standard deviation. The reference parameter values correspond to the choice provided in section~\ref{sec:reference_param_stem}.}
\label{fig:stem_cells_steady}
\end{figure}

\newpage
